\documentclass[moor]{informs1}              

\usepackage{natbib}
 \NatBibNumeric
 \def\bibfont{\small}%
 \bibpunct[, ]{[}{]}{,}{n}{}{,}%

\usepackage[colorlinks=true,breaklinks=true,bookmarks=true,urlcolor=blue,
     citecolor=blue,linkcolor=blue,bookmarksopen=false,draft=false]{hyperref}

\def\EMAIL#1{\href{mailto:#1}{#1}}
\def\URL#1{\href{#1}{#1}}         

\TheoremsNumberedThrough     

\EquationsNumberedThrough    

\usepackage{setspace}
\usepackage{graphicx}
\usepackage[dvipsnames]{xcolor}
\usepackage{xcolor}
\usepackage{environ}
\usepackage{tikz}
\usepackage{authblk, todonotes}
\usepackage{bbm}
\usepackage{nth}
\usepackage{dsfont,amsfonts}
\usepackage{subfig}
\usepackage{algorithm2e} 
\usepackage[noabbrev,nameinlink,capitalize,sort]{cleveref}
\usepackage{booktabs}
\usepackage[a4paper, total={5.7in, 9in}]{geometry}
\usepackage{enumerate, enumitem}
\usepackage{wrapfig}
\usepackage{pifont}
\newcommand{\xmark}{\ding{55}}

\usepackage{tikz}
\usepackage{tikzsymbols}
\usepackage{xcolor}

\definecolor{arm}{RGB}{100,140,171}
\usetikzlibrary{shapes.geometric}
\usetikzlibrary{shapes,arrows}
\usetikzlibrary{positioning}
\tikzstyle{theta}=[shape=circle,draw=mgreen,fill=mgreen!10]
\tikzstyle{state}=[shape=circle,draw=mred,fill=mred!10]
\tikzstyle{hidden}=[shape=circle,draw=mblue,fill=mblue!10]
\tikzstyle{gru}=[shape=rectangle,draw=black!50,fill=lime!20]
\tikzstyle{FFN}=[shape=rectangle,draw=black!50,fill=lime!20]
\tikzstyle{obs}=[shape=circle,draw=mblue,fill=mblue!10]
\tikzstyle{lightedge}=[<-,dotted]
\tikzstyle{mainstate}=[state,thick]
\tikzstyle{mainedge}=[<-,thick]	

\definecolor{mgreen}{rgb}{0,0.455,0.247}
\definecolor{mblue}{rgb}{0.098,0.18,0.357}
\definecolor{mred}{rgb}{0.902,0.4157,0.0196}
\definecolor{mgrey}{rgb}{0.90196,0.90,0.90}
\definecolor{mblack}{rgb}{0,0,0}

\newtheorem{property}{Properties}[section]

\newcommand{\comp}{{\complement}}
\newcommand{\tc}{{t.c.}}

\newcommand{\nn}{weak recursive}

\newcommand{\N}{{\mathbb{N}}}
\newcommand{\E}{{\mathbb{E}}}
\newcommand{\Q}{{\mathbb{Q}}}
\renewcommand{\P}{{\mathbb{P}}}
\newcommand{\q}{{\mathcal{Q}}}

\newcommand{\R}{{\mathds{R}}}

\newcommand{\mF}{{\mathcal{F}}}
\newcommand{\A}{{A^{\rho}_t}}

\newcommand{\mT}{{\mathcal{T}}}
\newcommand{\mbT}{{\overline{\mathcal{T}}}}
\newcommand{\tp}{{\{\rho_t\}_{t\in \mT}}}

\newcommand{\tT}{{t \in \mT}}
\newcommand{\tbT}{{t \in \mbT}}

\newcommand{\p}[1]{\rho_{#1}}
\newcommand{\lp}[1]{L^\infty_{#1}}

\DeclareMathOperator*{\esssup}{ess\,sup}
\DeclareMathOperator*{\essinf}{ess\,inf}

\newcommand{\ep}{{\varepsilon}}
\renewcommand{\vs}{{\varsigma}}

\newcommand{\Uset}{consolidated uncertainty set}
\renewcommand{\u}{{\mathfrak{u}}}
\newcommand{\U}{{\mathfrak{U}}}

\newcommand{\cha}[1]{#1}

\begin{document}

\VOLUME{}%
\NO{}%
\MONTH{}
\YEAR{}
\FIRSTPAGE{}%
\LASTPAGE{}%
\SHORTYEAR{}
\ISSUE{} %
\LONGFIRSTPAGE{} %
\DOI{}%

\RUNAUTHOR{Moresco, Mailhot, and Pesenti}

\RUNTITLE{Uncertainty Propagation and Dynamic Robust Risk Measures}

\TITLE{Uncertainty Propagation and Dynamic Robust Risk Measures}

\ARTICLEAUTHORS{%
\AUTHOR{Marlon R. Moresco}
\AFF{Department of Mathematics and Statistics, Concordia University, \EMAIL{marlonmoresco@hotmail.com}, \URL{}}
\AUTHOR{Mélina Mailhot}
\AFF{Department of Mathematics and Statistics, Concordia University, \EMAIL{melina.mailhot@concordia.ca}, \URL{}}
\AUTHOR{Silvana M. Pesenti}
\AFF{Department of Statistical Sciences, University of Toronto, \EMAIL{silvana.pesenti@utoronto.ca}, \URL{}}

\vspace{2em}\small\today
} 

\ABSTRACT{%
We introduce a framework for quantifying propagation of uncertainty arising in a dynamic setting. Specifically, we define dynamic uncertainty sets designed explicitly for discrete stochastic processes over a finite time horizon. These dynamic uncertainty sets capture the uncertainty surrounding stochastic processes and models, accounting for factors such as distributional ambiguity. Examples of uncertainty sets include those induced by the Wasserstein distance and $f$-divergences. 

We further define dynamic robust risk measures as the supremum of all candidates' risks within the uncertainty set. In an axiomatic way, we discuss conditions on the uncertainty sets that lead to well-known properties of dynamic robust risk measures, such as convexity and coherence. Furthermore, we discuss the necessary and sufficient properties of dynamic uncertainty sets that lead to time-consistencies of dynamic robust risk measures. We find that uncertainty sets stemming from $f$-divergences lead to strong time-consistency while the Wasserstein distance results in a new time-consistent notion of \cha{weak recursiveness}. Moreover, we show that a dynamic robust risk measure is strong \cha{time-consistent} or \cha{weak recursive} if and only if it admits a recursive representation of one-step conditional robust risk measures arising from static uncertainty sets.
}%

\KEYWORDS{Dynamic Risk Measures, Time-consistency, Distributional Uncertainty, Wasserstein distance}

\maketitle

\section{Introduction.}
As uncertainty prevents perfect information from being attained, decision makers are confronted with the consequences of their risk assessments made under partial information. Incorporating model misspecification and Knightian uncertainty into dynamic decision making, thus robustifying one's decisions, has been studied in various fields, including economics \cite{epstein2003recursive,Maccheroni2006JEF,Siniscalchi2011TE}, mathematical finance \cite{Bielecki2022WP,Neufeld2023MF}, and risk management \cite{acciaio2012risk}. Many circumstances require sequential decisions, where risk assessments are made over a finite time horizon and are based on the flow of information. Importantly, these decisions need to be time-consistent \cha{(t.c.)} and account for the propagation of uncertainty. 
Although the theory of \tc\, dynamic risk measures is growing \cite{riedel2004dynamic,cheridito2006dynamic,artzner2007coherent,jobert2008valuations,cheridito2011composition,bielecki2018unified,Feinstein2022SRM}, the \cha{incorporation of dynamic} uncertainty \cha{to dynamic risk measures is only} little explored. \cha{In the economic literature, the theory of recursive multiple-priors and variational preferences are closest to our work. While working with a recursive notion of time-consistency, most works focus on dynamic utility, event trees, and uncertainty sets that are fixed throughout time see e.g., \cite{epstein2003recursive,Maccheroni2006JEF, Siniscalchi2011TE} and references therein. Here, we work with dynamic risk measures, stochastic processes, and uncertainty that may change over time. Another related area of research is reinforcement learning (RL). While there are recent works in the field of distributional robust RL in the context of Markov Decision processes that account for uncertainty in the underlying processes, they typically maximise expected reward. The distributional robust risk-aware RL literature, which instead of expected rewards optimises risk measures, mostly deals with static uncertainty and static risk measures. Indicatively
see e.g, \cite{Abdullah2019WP, Smirnova2019WP} who model uncertainty on the transition probabilities, and \cite{Jaimungal2022SIAM} who consider uncertainty only at terminal time.}

In this work, we propose an axiomatic framework for quantifying uncertainty of discrete time stochastic processes \cha{and tie them to robustified dynamic risk measures}. Specifically, we introduce \emph{dynamic uncertainty sets} consisting of a family of time-$t$ uncertainty sets. Each time-$t$ uncertainty set is a collection of $\mF_t$-measurable random variables summarising the uncertainty of the entire stochastic process at time $t$. The dynamic uncertainty sets may vary with each stochastic process, as the uncertainty of two processes may differ, even if they share the same law. This general framework includes, to the authors knowledge, all uncertainty sets encountered in the literature, from moment constraints, $f$-divergences, semi-norms, the popular (adapted) Wasserstein distance, \cha{and ambiguity in a base probability}. 

Equipped with a \cha{strong t.c.} dynamic risk measure and a dynamic uncertainty set, we  define \emph{dynamic robust risk measures} as sequences of conditional robust risk measures, by taking the supremum of all risks in the uncertainty set. We then proceed by studying conditions on the dynamic uncertainty set that lead to well-known properties of dynamic robust risk measures such as convexity and coherence. Crucial to the dynamical framework are notions of time-consistencies, of which many have been introduced and studied in the literature. The most common is strong time-consistency \cha{-- often also referred to as recursiveness --}, leading to a dynamic programming principle \cite{cheridito2006dynamic, ruszczynski2010risk, Cheng2022WP}. While the majority of works assume normalisation of dynamic risk measures, in a robust setting, incorporating uncertainty does not necessarily \cha{result in the robustified dynamic risk measures being normalised}. Indeed, an important subject of debate is whether the value of zero -- or more generally an $\mathcal{F}_{t-1}$-measurable random variable -- contains uncertainty -- at time $t$. We find that uncertainty sets induced by the $f$-divergence are normalised while those generated by the Wasserstein distance or a norm, \cha{with e.g. a constant tolerance distance}, are not. Consequently, we introduce a new concept of \cha{weak recursiveness} to account for uncertainty sets that are not normalised.

One of the manuscript's key theorems generalises results from the seminal works of \cite{cheridito2006dynamic,ruszczynski2010risk} \cha{to account for dynamic uncertainty}. Specifically, we show that a dynamic robust risk measure is strong \tc\, or \nn\, if and only if it admits a recursive representation of one-step robust risk measures. Furthermore, these one-step robust risk measures are characterised by dynamic uncertainty sets which possess the property of static. Static uncertainty sets arise in one-period settings and \cha{only account for uncertainty at the current time.} Thus, we show that when working with \cha{strong \tc\, or weak recursive} dynamic robust risk measures, it is enough to consider the simpler subclass of static uncertainty sets.

The paper is organised as follows. \Cref{DynamicSetup} presents the framework of dynamic uncertainty sets, discusses their properties, and provides examples. In  \Cref{DynamicRRM}, we define dynamic robust risk measures and show, in an axiomatic way, \cha{sufficient and necessary condition on}
the dynamic uncertainty set that provide properties of dynamic robust risk measures. \cref{sec: notion of tc} introduces different notions of time-consistencies of dynamic uncertainty sets and connects them to time-consistencies of dynamic robust risk measures. Furthermore, in \cref{sub-sec:construction-time} we \cha{prove the recursive representation of strong t.c. and weakly recursive} dynamic robust risk measures. We conclude with examples of  dynamic robust risk measures in \cref{sec:example-rm}. Proofs not included in the main body are delegated to Appendix \ref{app:proofs}.

\section{Uncertainty in a Dynamic Setting.}
\label{DynamicSetup}
This section proposes a framework for axiomatically quantifying uncertainty in a dynamic setting. 

\subsection{Preliminaries and Notation.}
Consider a finite time horizon $T \in \N$, and a filtered probability space $(\Omega,\mF,\P, \{\mF_t\}_{t \in \cha{\{0, 1, \ldots, T\}}})$. For $t \in \cha{\{0, 1, \ldots, T\}}$, we denote by $L^\infty_t := L^\infty(\Omega,\mF_t,\P)$ the space of $\mF_t$-measurable bounded random variables and set $L^\infty_{0,T} := L^\infty_0 \times \ldots \times L^\infty_T$. This setup allows for the representation of one-dimensional stochastic processes $X \in L^\infty_{0,T}$, $X := \{ X_0, X_1, \ldots , X_T \}$, where each $X_t \in L^\infty_t$ represents a (discounted) loss at time $t$, \cha{and if not otherwise stated we set for simplicity $X_0 = 0$}. We further define the spaces $L^\infty_{t,s} :=  0\times \ldots \times 0 \times L^\infty_t \times \ldots \times L^\infty_s \times 0\times  \ldots \times 0 $, for all $0\leq t<s \leq T$. For $X \in L^\infty_{0,T} $, we denote by $X_{t:s}$, $0\leq t<s \leq T$, its projection onto 
the space $L^\infty_{t,s}$, that is $X_{t:s} := \{0, \ldots,0,X_t,\ldots,X_s,0,\ldots,0\} $. Whenever there is no confusion, we omit the $0$'s, i.e. we write $X_{t:s} = \{ X_t, \ldots, X_s\} $. Thus, for $X_{s:T}\in\lp{s,T}$ and $Z_t$ with $t<s$, we may write $X_{s:T} + Z_t =(\underbrace{0,\dots,0}_{t},Z_t,\underbrace{0,\dots,0}_{s-t-1} , X_s,\dots, X_T)$.
We further define the supremum norm \cha{on} the spaces $\lp{t,s}$ as
\begin{equation}\label{def:norm}
    \Vert X_{t:s} \Vert_{t,s} 
    :=  \essinf\left\{ m \in \lp{t} : \sup_{t\leq i \leq s} | X_i| \leq m   \right\} \,.
\end{equation}
If not otherwise stated, all equalities and inequalities between random vectors are component-wise and in a $\P$-almost sure (a.s.) sense.
Central to the exposition are set-valued functionals. To clarify the notation, we recall that the sum of sets, 
\begin{align*}
    A + B  := \left\{ X + Y \in \lp{0,T} : X \in A, \, Y \in B  \right\}
    \,,
    \quad \;\text{where}\quad A,B \subseteq \lp{0,T}\,.
   \end{align*}
By abuse of notation, we may denote sets consisting of a singleton by its element, i.e., $Z:= \{Z\}\subset \lp{0,T}$. We further recall the multiplication of a set $A \subseteq \lp{t}$ with a $\mF_t$-measurable random variable $\lambda \in \lp{t}$ as
\begin{equation*}
     \lambda \, A := \big\{ \lambda \, X \in \lp{t}  :X \in A  \big\}\,,
\end{equation*}
and denote the complement of a set $A \subseteq \lp{t}$ by $A^\complement := \{X \in \lp{t} : X \notin A   \}$.

\subsection{Dynamic Uncertainty Sets.}

In this section, we introduce the notion of \emph{dynamic uncertainty sets}, that quantifies uncertainty around stochastic processes. \cha{For this, we define the notation $\mT := \{0, \ldots, T-1\}$ and $\mbT := \{1, \ldots, T\}$.}

\begin{definition}[Dynamic Uncertainty Set]
A dynamic uncertainty set $\u: = \{u_t\}_{t \in \cha{\mbT}}$  is a sequence of time-$t$ uncertainty sets $\{u_t\}_{t \in \cha{\mbT} } $, where for each $t \in \cha{\mbT}$, the time-$t$ uncertainty set $u_t$ is a mapping $u_t : L^\infty_{0,T} \rightarrow 2^{\lp{t}}$.
\end{definition}

A time-$t$ uncertainty set $u_t$ is thus a set function mapping a stochastic process $X_{0:T}$ to a subset of $\mF_t$-measurable random variables and could thus include uncertainty of the entire time horizon of the processes. 
If an uncertainty set is  evaluated on a projection of a stochastic process, e.g., on $X_{t:s}$, $t,s\in \cha{\mbT}$ with $t\le s$, then we simply write 
$u_t (X_{t:s}) := u_t(0, \dots , 0,X_t, \dots , X_s , 0, \dots, 0 ) \subseteq \lp{t}$.

\cha{
In the context of, e.g., financial losses represented by $X_{t:T}$, uncertainty at time $t$ may arise as the agent encounters ambiguity whether $X_{t:T}$ accurately models the true loss. In such instances, it becomes prudent for the agent to consider a set of alternative stochastic processes. These alternative stochastic process are losses that are considered ``close'' to $X_{t:T}$ or share common attributes such as similar distributional features. The uncertainty at time $t$, i.e. $u_t(X_{t:T})$, may then be viewed as the projection of the space of alternative processes onto time $t$. 
  In \cref{sec:examples}  we present several examples of uncertainty sets from in the literature adapted to the dynamic setting.}

\begin{property}\label{property-u}
A time-$t$ uncertainty set $u_t$ may satisfy the following 
\begin{enumerate}[label=\roman*)]

    \item \textbf{Proper:}\label{property:proper}
    Non-empty and bounded from above\footnote{A set $u \subset \lp{t} $ is bounded  if $\inf \{ c \in \R : \P( c \geq |X|) =1, \forall \; X \in u   \}  < \infty $, and bounded from above if $\inf \{ c \in \R : \P( c \geq X) =1, \forall \; X \in u   \}  < \infty $.} for all $X_{t:T} \in \lp{t,T}$.    
    
    \item \textbf{Normalisation:}\label{property:norm}  $ u_t (0) := u_t(0,\dots ,0) = \{0\}$.

    \item \textbf{Order preservation:}\label{property:order}
    Let $X_{t:T} \leq Y_{t:T}$ with $X_{t:T}, Y_{t:T} \in\lp{t,T}$. Then
    for each $Z \in u_t(X_{t:T}) $ there exists a $W \in u_t(Y_{t:T})$ such that $Z\leq W$.

    \item \textbf{Monotonicity:}\label{property:mon}   $X_{t:T} \leq Y_{t:T}$ implies that $u_t (X_{t:T}) \subseteq u_t(Y_{t:T})$, for all $X_{t:T}, Y_{t:T} \in \lp{t,T}$.

    \item \textbf{Translation invariance:}\label{property:trans}
    $ u_t(X_{t:T}+Z_s) =  u_t(X_{t:T})  + Z_s$ for all $X_{t:T} \in \lp{t,T}$ and $ Z_s \in L^\infty_{s}$ with $s<t$.\footnote{Recall that $u_t (X_{t:T}) + Z_t := \big\{ Y + Z_t \in \lp{t} : Y \in  u_t (X_{t:T}) \big\}$.}

    \item \textbf{Static:}\label{property:static}
    $u_t(X_{t:T}) = u_t(X_t)$, for all $X_{t:T} \in \lp{t,T}$.

     \item \textbf{Locality:}\label{property:local}
     $u_t (1_B \, X_{t:T} + 1_{B^\comp} \, Y_{t:T}) = 1_B \, u_t(X_{t:T}) + 1_{B^\comp} \, u_t(Y_{t:T})$ for all  $B \subseteq \lp{t}$ and $X_{t:T}, Y_{t:T} \in\lp{t,T}$.
    
    \item \textbf{Positive homogeneity:}\label{property:pos-hom}
    $u_t (\lambda \, X_{t:T} ) = \lambda \,u_t ( X_{t:T} ) + 1_{\lambda = 0 }\, u_t (0 )  $ for all $0 \leq \lambda \in \lp{t-1}$ and $X_{t:T} \in\lp{t,T}$.

    \item \textbf{Star-shapedness:}\label{property:star}
    $u_t (\lambda\, X_{t:T} ) \subseteq \lambda \, u_t ( X_{t:T} ) + 1_{\lambda = 0 }\, u_t (0 )  $ for all $ \lambda \in \lp{t-1} $ with $0 \leq \lambda \leq 1$ and $X_{t:T} \in\lp{t,T}$.

\end{enumerate}
\end{property}

We say a dynamic uncertainty set $\u = \{u_t \}_{t\in\cha{\mbT}}$ satisfies one of the above properties if  $u_{t} $ satisfies it for all $t \in\cha{\mbT}$. 
Clearly, a time-$t$ uncertainty set should be non-empty and bounded. Normalisation pertains to whether $0$ is ambiguous, a property that may or may not be desired. In \cref{sec:examples} we see that uncertainty sets induced by norms are in general not normalised while those by $f$-divergences are. \cha{In the context of hedging, an optimal strategy may attain a liability of $X_{0:T} = 0$. However, the underlying model may not capture all aspects, such as market liquidity and legal  and political risk, which can be accounted for through a non-normalised uncertainty set.}
Monotonicity, which states that the uncertainty set of a dominating stochastic process is larger than the original process, is a strong property, however important for the exposition. The weaker notion of order preservation (implied by monotonicity), states that any element in $u_t(X_{t:T})$ is \cha{a.s.} dominated by an element in the uncertainty set of the dominating process $Y_{t:T}$. 
Translation invariance means that known information does not change the uncertainty set and is useful for incorporating prior information, such as Bayesian updating. 
For example, if $\u$ is normalised and translation invariant, then any $\mF_{t-1}$-measurable random variable does not exhibit uncertainty at time $t$.
A static \cha{time-$t$} uncertainty set \cha{only accounts for uncertainty around $X_t$. Indeed}, if $\u$ is static then it satisfies $u_t(X_{t}) = u_t(X_{t} + Y_{t+1:T}) $  for any $Y_{t+1:T} \in \lp{t+1,T}$, thus the uncertainty sets is indifferent about the future of the process. 
Positive homogeneity implies that the uncertainty set scales with the size of the loss, while star-shapedness means that it increases with size. 

\begin{lemma}\label{lemma-u-properties}
Let $u_t$ be a time-$t$ uncertainty set. Then, 
\begin{enumerate}[label = \roman*)]
    \item \label{lemma-u-properties-1}
    $u_t$ is normalised and local if and only if  $u_t \left(1_B \, X_{t:T}\right) = 1_B \, u_t\left(X_{t:T}\right)$ for all $B \in \mF_{t-1}$ and $X_{t:T} \in \lp{t,T}\,$;
    \item \label{lemma-u-properties-2}
    if $u_t$ is positive homogeneous, then $u_t$ is local.
\end{enumerate}
\end{lemma}
The proof is delegated to Appendix \ref{app:proofs}.

\subsection{Examples of Dynamic Uncertainty Sets.}\label{sec:examples}
\cha{In the static setting, there is a plethora of uncertainty sets considered in literature, including} uncertainty sets defined via e.g., mixtures of distributions \cite{zhu2009worst}, moment constraints \cite{Hurlimann2002ASTIN,natarajan2009constructing}, divergence constraints \cite{Jaimungal2022SIAM,luo2020distributionally,ben2013robust,calafiore2007ambiguous,wang2016likelihood}, and combinations of moment and divergence constraints \cite{bernard2022robust}. \cha{Furthermore, in the study of convex risk measures, robustness is closely associated with the underlying probability measure, with \cite{artzner1999coherent,delbaen2006structure} emphasising uncertainty sets centred around a probability measure. However, a limitation of the latter perspective is that a random variable's uncertainty only depends on its distribution. The proposed approach goes beyond by considering uncertainty sets of random variables rather than probability measures. }

\cha{In this section we focus on examples of uncertainty sets that are} constructed as ``balls'' around a reference distribution with radius given by a tolerance distance. In the dynamic setting, the \cha{time-$t$} reference distribution \cha{could be} the distribution of \cha{$X_{t:T}$} conditional on $\mF_{t-1}$, to e.g., \cha{account for uncertainty in} transition probabilities. \cha{The time-$t$ tolerance distance may depend on $X_{t}$ and encompass information available at time $t$ such as market cyclicality, trading volume, and market liquidity. }
The first set of examples pertains to uncertainty sets induced by semi-norms or norms on the spaces of random variables and stochastic processes, such as the total variation norm, $p$-norm, H\"older norm, and supremum norm, see e.g., \cite{gotoh2011role,gotoh2013robust} for the static setting.
Second, we discuss uncertainty sets induced by the (adapted) Wasserstein distance. Uncertainty sets characterised by the Wasserstein distance in the static case have become popular, indicatively see \cite{pflug2007ambiguity,mohajerin2018data,gao2022distributionally,pesenti2020portfolio,Jaimungal2022SIAM}, and \cite{backhoff2020adapted} for the adapted Wasserstein distance. We also refer to \cite{blanchet2019quantifying} for uncertainty sets based on optimal transport between probability measures.
The last set of examples considers distances on the space of probability distributions, and in particular, we consider dynamic uncertainty sets induced by $f$-divergences and the Kullback-Leibler (KL) divergence. \cref{tab:u_set_properties} summarises the properties of the dynamic uncertainty sets. 

\begin{example}[Semi-Norm on Random Variables]\label{example: norms}
Consider the dynamic uncertainty set given by
\begin{equation}\label{eq:ex:u-seminorm}
    u_t^{||\cdot||}(X_t)
    := 
    \big\{
    Y \in \lp{t} ~:~
    \Vert X_t - Y \Vert \le \ep_{X_t}
    \big\}\,,
    \quad  \tbT\,,
\end{equation}
where $\Vert \cdot \Vert \colon \lp{t} \to \lp{t-1}$ is a (random) semi-norm and $ \ep_{X_t}\ge 0$, $\ep_{X_t} \in \lp{t-1}$, a tolerance distance. 
The choice of tolerance distance includes, for example, $\ep_{X_t} = \ep$\cha{, $\ep_{X_t} = (T-t)\ep$}, $\ep_{X_t}  =\ep \,\text{var}(X_t|X_{t-1})$, with $\ep\in\R$ and where $\text{var}(\cdot)$ denote the conditional variance. 

This time-$t$ uncertainty set is proper and static. It is normalised if and only if $\Vert \cdot \Vert$ is a norm and satisfies $\ep_{0} = 0$, that is, $\ep_{X_t} = 0$ whenever $X_t = 0$.  If $\ep_{X_t} \le \ep_{Y_t}$ for all $X_t \leq Y_t\in\lp{t}$, then the uncertainty set is order preserving. 
It is translation invariant, if $\ep_{X_t + Y_{t-1}} = \ep_{X_t}$ for all $X_t \in \lp{t}$ and $Y_{t-1} \in \lp{t-1}$. If $\ep_{\lambda X_t} = \lambda\ep_{X_t}$, for all $X_t \in \lp{t}$ and $0 \leq \lambda \in \lp{t-1}$, then the uncertainty set is  positive homogeneous, and consequently local. Similarly, if   $\ep_{\lambda X_t} \leq \lambda\ep_{X_t}$ for all $X_t \in \lp{t}$ and $0 \leq \lambda \in \lp{t-1}$, then the uncertainty set is star-shaped. 
Moreover, if  $ \ep_{ X_t + Y_t} \geq \ep_{X_t} + \ep_{Y_t}$ for all $X_t, Y_t \in \lp{t}$, then $ u_t( X_t) + u_t (Y_t) \subseteq u_t( X_t + Y_t)$\,.

For the special case when the tolerance distance  $\ep^\dagger \in \lp{t-1}$ satisfies $\ep^\dagger = \ep^\dagger_{X_t} $ for all $X_t\in \lp{t}$, then
the uncertainty set \eqref{eq:ex:u-seminorm} reduces to
\begin{align}\label{eq:ex:u-seminorm-spacial}
u_t^{||\cdot||}\big( X_t\big) 
 &= 
    \{Y +X_{t} \in \lp{t} : \Vert Y    \Vert \leq \ep^\dagger \}
=  u_t^{||\cdot||}(0) + X_t\,,
\end{align}
which implies that the uncertainty set of $X_t$ is entirely described by the uncertainty around the origin.
\end{example}

\begin{example}[Semi-Norm on Stochastic Processes]\label{example: norms of t:T}
Consider the dynamic uncertainty set with a semi-norm $\cha{\Vert \cdot \Vert}\colon \lp{t,T} \to \lp{t}$ and tolerance distance $0<\ep_{X_{t:T}}\in\lp{t}$, given by   \begin{equation}\label{eq:uncertainty-ex-main}
    u_t(X_{t:T}) :=    \big\{    Y \in \lp{t} ~:~    \cha{\Vert} X_{t:T}- Y \cha{\Vert} \le \ep_{X_{t:T}}    \big\}\,,
    \quad \tbT\,.
\end{equation}
If the norm is the supremum norm given in \eqref{def:norm}, i.e., $ \cha{\Vert} \cdot \cha{\Vert} := \Vert \cdot\Vert_{t:T}$, then the uncertainty set becomes, \cha{ for $t<T$}
\begin{equation*}
    u_t(X_{t:T}) =    \Big\{    Y \in \lp{t} ~:~     \max \big\{ \Vert  X_{t+1:T} \Vert_{t+1:T}\, , \;\esssup \{X_t-Y\}\big\} \le \ep_{X_{t:T}}    \Big\}\,.
\end{equation*}
If further $ \ep_{X_{t:T}} <   \Vert  X_{t+1:T} \Vert_{t+1:T}$, then $u_t(X_{t:T}) = \emptyset$, and if  $ \ep_{X_{t:T}} \ge   \Vert  X_{t+1:T} \Vert_{t+1:T}$, then $u_t$ simplifies to
\begin{equation}\label{eq:uncertainty-ex}
    u_t(X_{t:T}) :=    \big\{    Y \in \lp{t} ~:~    \Vert X_{t}- Y\Vert_{t} \le \ep_{X_{t:T}}    \big\}\,.
\end{equation}

When $ \cha{\Vert} \cdot  \cha{\Vert}$ is the sum of norms, i.e., $\cha{\Vert} Y_{t:T}- X_{t:T}\cha{\Vert} := \sum_{s = t}^T\Vert X_s-Y_s \Vert_{s:s}$, and $ \ep_{X_{t:T}}  < \ep_{X_t} +  \cha{\Vert} X_{t+1:T} \cha{\Vert}$, for some $\ep_{X_t}>0$, then the uncertainty set is empty. For $\ep_{X_{t:T}} = \ep_{X_t} +  \cha{\Vert} X_{t+1:T} \cha{\Vert}$, $u_t$ becomes identical to the one in \cref{example: norms}.

Alternatively, we can define the uncertainty set as the $\lp{t}$-projection of stochastic processes, that is
   \begin{align*}
           u_t(X_{t:T}) &:=    \big\{    Y_t \in \lp{t} ~:~     \cha{\Vert} X_{t:T}- Y_{t:T} \cha{\Vert} \le \ep_{X_{t:T}}    \big\} \,.
    \end{align*}     
For the H\"older and total variation norm, this uncertainty set is equal to the entire space of $\mF_t$-measurable random variables, i.e., $u_t(X_{t:T}) = \lp{t}$, and thus is not proper. For the $p$-norm, the uncertainty set reduces to \eqref{eq:uncertainty-ex} with the $p$-norm on $\lp{t}$. Thus, \cha{many uncertainty sets based on} norms for processes lead to pathological (not proper) uncertainty sets or reduce to those in \cref{example: norms}.
\end{example}

\begin{example}[Wasserstein Uncertainty]\label{ex:wasserstein-u}
Consider the time-$t$ uncertainty set induced by the $p$-conditional Wasserstein distance, $ p \ge 1$,
\begin{equation}\label{eq: wasserstein u set}
     u_t^W(X_{t}) :=    \left\{    Y \in \lp{t} ~:~   \int_0^1 | F_{Y|\mF_{t-1} }^{-1} (\alpha) - F_{X_t|\mF_{t-1}}^{-1} (\alpha) |^p d\alpha \leq \ep_{X_t}^p         \right\}\, ,
\end{equation}
where $F_{Y|\mF_{t-1} }^{-1}$ denotes the \cha{conditional} left-quantile function of $Y$ given $\mF_{t-1}$\cha{, see e.g. \cite{de2023conditional} for a definition.}
This uncertainty set is order-preserving, translation invariant, positive homogeneous, and normalised if $\ep_{X_t}$ satisfies the respective properties as in \cref{example: norms}. \cha{Uncertainty sets of this type are studied in \cite{esfahani2015data,luo2019decomposition,luo2020distributionally,gao2022distributionally}, where $\ep_{X_t}$ is not a function of $X_{t}$ and therefore, the uncertainty sets are not normalised.}

Alternatively, the Wasserstein distance for stochastic processes gives raise to  
\begin{align*}
     u_t(X_{t:T}) &:=    \left\{    Y_t \in \lp{t} :  W(X_{t:T},Y_{t:T})  \leq \ep_{X_{t:T}}        \right\} \,, \quad \text{with}
\\ W(X_{t:T},Y_{t:T}) &:=  \inf \{  (\Vert X'_{t:T} - Y'_{t:T} \Vert^p )  : F_{X'_{t:T}|\mF_{t-1}} = F_{X_{t:T}|\mF_{t-1}} ,\; F_{Y'_{t:T}|\mF_{t-1}} = F_{Y_{t:T}|\mF_{t-1}} \}\,,
     \end{align*}
 where the infimum is taken over all joint distributions $(X'_{t:T}, Y'_{t:T})$ with given conditional marginals, $F_{Y|\mF_{t-1}}$ denotes the conditional cumulative distribution function (cdf) of $Y$ given $\mF_{t-1}$, and $\Vert \cdot \Vert^p$ is the $p$-norm. This uncertainty set has the same properties as the one in \eqref{eq: wasserstein u set}. Since $\Vert X'_{t} - Y'_{t} \Vert^p \leq \Vert X'_{t:T} - Y'_{t:T}  \Vert^p   \leq \Vert X'_{t} - Y'_{t} \Vert^p +\Vert X'_{t+1:T} - Y'_{t+1:T}  \Vert^p $ and we can choose $Y'_{t+1:T} = X'_{t+1:T} $, it follows that $\Vert X'_{t} - Y'_{t} \Vert^p = \Vert X'_{t:T} - Y'_{t:T}  \Vert^p$. Hence, if $\ep_{X_{t:T}} = \ep_{X_t} $, this uncertainty set reduces to \eqref{eq: wasserstein u set}.
The same \cha{holds} for the adapted Wasserstein distance. 
\end{example}

\begin{example}[Uncertainty on the probability]\label{ex: uncert on prob}
Uncertainty may arise from the underlying probability measure, such as in the context of model risk. For this, we denote by $F_X$ the cdf of  $X$ under the base probability measure $\P$. 
Further, let $\q$ be a set of probability measures that are absolutely continuous with respect to (w.r.t.) $\P$ and consider the uncertainty set
\begin{equation}\label{eq: u set probabilities}
    u_t^\q(X_{t}) := \left\{ Y  \in \lp{t}: F_{Y|\mF_{t-1}} = F_{X_t|\mF_{t-1}}^\Q, \text{ for some }  \Q \in \q \right\}\,,
\end{equation}
where $F^\Q_X$ is the cdf of $X$ under $\Q$. Then $u_t^\q$
is proper, normalised, order-preserving, translation invariant, local, and positive homogeneous. Such an uncertainty set is proposed in \cite{blanchet2019quantifying}, where $\q$ is based on optimal transport between probability measures. \cha{
When $\q$ contains probability measures that are not absolutely continuous w.r.t. $\P$, the above properties of $u_t^\q$ may not hold. To illustrate, let $\P,\Q \in \q$ where $\Q$ is not absolutely continuous w.r.t. $\P$. Then there exists a set $A$ such that $\P(A) = 0 < \Q(A)$. Consequently, as $1_A = 0$ $ \P$-a.s., we have $u^\q_1 (0) = u^\q_1 (1_A) \supset \{0\}$, where the inclusion is strict.   }
\end{example}

\begin{example}[Uncertainty induced by Divergences]\label{ex: uncer divergence}
    Let $D_t$ be a function  mapping cdfs to $\lp{t-1}$, i.e.  $(F,G) \mapsto D_t(F,G) \in \lp{t-1}$, and consider its induced uncertainty set \begin{equation}\label{eq: u set distribution}
    u_t^D(X_{t:T}) := \big\{ Y_t  \in \lp{t}:   D_t (F_{Y_{t:T}},F_{X_{t:T}}) \leq \ep_{X_{t:T}}      \big\}\,.
\end{equation}
Examples of $D_t$ include $f$-divergences and, in particular, the KL divergence. The uncertainty set 
$u_t^D$ is normalised, whenever the divergence of distributions with differing support is equal to infinity -- which is the case for $f$-divergences. It is translation invariant, if $D_t (F_{Y_{t:T}},F_{X_{t:T}+c}) = D_t (F_{Y_{t:T}-c},F_{X_{t:T}})$, for all $c \in \lp{t-1}$, and $\ep_{X_{t:T}} = \ep_{X_{t:T}+c }$. It is positive homogeneous, if  $D_t (F_{Y_{t:T}},F_{\lambda X_{t:T}}) = D_t (F_{\frac{Y_{t:T}}{\lambda}},F_{X_{t:T}})$ and $\ep_{X_{t:T}}=\ep_{\lambda X_{t:T}} $, for all $0<\lambda \in \lp{t-1}$.
Uncertainty sets induced by conditional $f$-divergences, in particular, the conditional KL-divergence, satisfy the above. Furthermore, uncertainty  sets induced by $f$-divergences are order preserving if $\ep_{X_{t:T}} \le \ep_{Y_{t:T}}$ for all $X_{t:T} \le Y_{t:T}$.

If $\q$ is given by $\q := \{ \Q : d_t (F_{X_{t:T}},F^\Q_{X_{t:T}}) \leq \ep    \}$, then the uncertainty sets in Equations \eqref{eq: u set probabilities} and \eqref{eq: u set distribution} coincide. 
\end{example}

\begin{table}[h]
 \centering
 \begin{tabular}{l@{\hskip 0.2in} c@{\hskip 0.25in} c@{\hskip 0.25in} c@{\hskip 0.25in} c}
    Properties & Semi-norm   & 
Wasserstein &  
Probability  & cond. KL\\[0.5em]
\toprule[1pt] \toprule[1pt] \\[-0.5em]
& Eq. \eqref{eq:ex:u-seminorm}, $u_t^{||\cdot||}$  & \;Eq. \eqref{eq: wasserstein u set}, $u_t^{W}$ &  Eq.
\eqref{eq: u set probabilities}, $u_t^{\q}$
& Eq. \eqref{eq: u set distribution}, $u_t^{KL}$\\[0.5em]
    \toprule[1pt]\\[-.7em]
    proper & \checkmark & \checkmark & \checkmark & \checkmark\\
    normalised & norm, $\ep_0 = 0$ & $\ep_0 = 0$  & \checkmark & \checkmark\\
    order preserving & $\ep_{X_t} \le \ep_{Y_t} $ & $\ep_{X_t} \le \ep_{Y_t} $ & \checkmark & $\ep_{X_t} \le \ep_{Y_t} $\\
    translation invariant & $\ep_{X_t + Y_t} = \ep_{X_t}  $ & $\ep_{X_t + Y_t} = \ep_{X_t}  $ & \checkmark & \checkmark\\ 
    static & \checkmark & \checkmark & \checkmark & \xmark\\
    local & $ \ep_{\lambda X_t} =\lambda\, \ep_{X_t}  $ & $ \ep_{\lambda X_t} =\lambda\, \ep_{X_t}  $ & \checkmark & \checkmark\\
    positive homogeneous &  $ \ep_{\lambda X_t} = \lambda\, \ep_{X_t}  $ &  $ \ep_{\lambda X_t} =\lambda\, \ep_{X_t}  $ & \checkmark & \checkmark\\[0.5em] 
    \toprule[1pt] \\[-0.5em]
\end{tabular}
    \caption{Examples of dynamic uncertainty sets \cha{discussed in \Cref{sec:examples}} and their properties.}
    \label{tab:u_set_properties}
\end{table}

\section{Dynamic Robust Risk Measure.}
\label{DynamicRRM}
Next, we \cha{propose a class of} dynamic robust risk measures that incorporates the dynamic uncertainty sets introduced in the last section. \cha{Specifically, we are interested in robustifying strong time-consistent (\tc)\, dynamic risk measures, that are normalised, monotone, and translation invariant, by taking at each time point the worst-case value of the dynamic risk measure. Strong \tc\, dynamic risk measures that are normalised, monotone, and translation invariant are studied extensively, as they allow for a recursive representation (see \Cref{thm:recursion}), that in many settings lead to a Dynamic Programming Principle \cite{ruszczynski2010risk} which allows to solve multi-step optimisation problems \cite{shapiro2021lectures}. 
When robustifying a dynamic risk measure, however, some of their characteristics may get lost, thus this section studies the necessary and sufficient requirements on dynamic uncertainty sets that preserves the properties of dynamic robust risk measures.}

\subsection{{\color{black}{Definition and Properties.}}}
We first recall the definitions and properties of conditional and dynamic risk measures and refer the reader to \cite{cheridito2004coherent,frittelli2006risk,follmer2016stochastic,ruszczynski2006conditional} for discussions and interpretations, as well as \cite{laeven2023dynamic} for star-shaped dynamic risk measures. 

\cha{As uncertainty may change over time, we consider the dynamic risk of the entire process rather than the total loss amount at terminal time. In particular, when incorporating uncertainty, we do not assume that the uncertainty sets respect translation invariance, hence working with the entire process becomes necessary.}

    \begin{definition}[Dynamic Risk Measure]\label{def-cond-rm}

    A dynamic risk measure on $\mT$ is a sequence of conditional risk measures 
    $ \{\p{t,T}\}_{ t\in \mT}$, where for each $t < s, \cha{\,t\in \mT ,s \in \mbT}$, the conditional risk measure $\rho_{t,s}$ is a mapping $\rho_{t,s} : L^\infty_{t+1,s} \to L^\infty_t$.
    \end{definition}

Thus, \cha{$\rho_{t,s}$ associates each stochastic process in $\lp{t+1,s}$ with a $\mF_t$-measurable random variable}. Whenever we write $\rho_{t,s}$ we implicitly assume that $t< s$ with $\cha{\,t\in \mT ,s \in \mbT}$.

\begin{property}\label{properties-rm}
A conditional risk measure $\rho_{t,s}$ may satisfy the following properties:   
\begin{enumerate}
         \item \textbf{Normalisation:} $\rho_{t,s} (0):= \rho_{t,s} (0, \ldots, 0)=0$.

        \item \textbf{Monotonicity:} $\rho_{t,s} (X_{t+1:s}) \leq \rho_{t,s} (Y_{t+1:s})$, for all $X_{t+1:s}, Y_{t+1:s} \in \lp{t+1,s}$ with $X_{t+1:s} \leq Y_{t+1:s}$.

        \item \textbf{Translation Invariance:} 
        $\rho_{t,s}\left( X_{t+1:s} + Y_t  \right)  =\rho_{t,s}( X_{t+1:s}) + Y_t $, for all 
        $X_{t+1:s} \in L^\infty_{t+1,s} $ and $Y_t \in \lp{t}$.

        \item \textbf{Locality:}
        $\rho_{t,s} (1_B\, X_{t+1:s} + 1_{B^\comp}\, Y_{t+1:s}) = 1_B \, \rho_{t,s}(X_{t+1:s}) +1_{B^\comp}\, \rho_{t,s}(Y_{t+1:s}) $, for all $X_{t+1:s},Y_{t+1:s} \in \lp{t+1,s}$ and $B$ that are $\mF_t$-measurable.

        \item \textbf{Positive Homogeneity:}
        $\rho_{t,s}( \lambda \, X_{t+1:s}) = \lambda \,\rho_{t,s}(X_{t+1:s})$, for all  $X_{t+1:s} \in\lp{t+1,s} \,$, and $\lambda \in \lp{t}$ with $\lambda \ge 0$.

        \item \textbf{Convexity:}
        $\rho_{t,s}(\lambda \, X_{t+1:s} + (1-\lambda )\, Y_{t+1:s} ) \leq \lambda \, \rho_{t,s} (X_{t+1:s}) + (1-\lambda) \, \rho_{t,s}(Y_{t+1:s})$, for all  $X_{t+1:s},Y_{t+1:s} \in \lp{t+1,s}$ and $\lambda \in \lp{t}$ with $0 \leq \lambda \leq 1$.

         \item \textbf{Sub-additivity:}
         $\rho_{t,s}( X_{t+1:s} + Y_{t+1:s} ) \leq  \rho_{t,s} (X_{t+1:s}) + \rho_{t,s}(Y_{t+1:s})$, for all $X_{t+1:s},Y_{t+1:s} \in \lp{t+1,s}$.

        \item \textbf{Concavity:}
        $\rho_{t,s}(\lambda\, X_{t+1:s} + (1-\lambda )\,Y_{t+1:s} ) \geq \lambda\,  \rho_{t,s} (X_{t+1:s}) + (1-\lambda) \, \rho_{t,s}(Y_{t+1:s})$, for all $X_{t+1:s},Y_{t+1:s} \in \lp{t+1,s}$ and $\lambda \in \lp{t}$ with $0 \leq \lambda \leq 1$.
        
        \item \textbf{Super-additivity:} 
        $\rho_{t,s}( X_{t+1:s} + Y_{t+1:s} ) \geq  \rho_{t,s} (X_{t+1:s}) + \rho_{t,s}(Y_{t+1:s})$, for all $X_{t+1:s},Y_{t+1:s} \in \lp{t+1,s}$.

        \item \textbf{Additivity:}  
        $\rho_{t,s}( X_{t+1:s} + Y_{t+1:s} ) =  \rho_{t,s} (X_{t+1:s}) + \rho_{t,s}(Y_{t+1:s})$, for all $X_{t+1:s},Y_{t+1:s} \in \lp{t+1,s}$.

        \item \textbf{Star-shapedness:} 
        $\rho_{t,s}(\lambda\,  X_{t+1:s}) \leq \lambda\, \rho_{t,s}(X_{t+1:s})$, for all $X_{t+1:s} \in \lp{t+1,s}$  and $\;\lambda \in \lp{t}$ with $0\leq \lambda \leq 1$.
    \end{enumerate}
\end{property}
We say a dynamic risk measure $\{\rho_{t,T} \}_{t\in\mT}$ satisfies one of the properties if  $\rho_{t,T} $ satisfies it for all $t \in\mT$. The conditional risk measure $\rho_{t,t+1}\colon L^\infty_{t+1} \to L^\infty_t$, $t \in \mT$, is called a one-step (conditional) risk measure and we denote it simply by $\rho_t$, i.e. $\rho_t (\cdot):= \rho_{t, t+1}(\cdot)$. 
The acceptance set of a conditional risk measure $\rho_{t,s}$ is defined by 
\begin{equation}\label{eq:def-acceptance-rm}
    A_{t,s}^{\rho} := \big\{X_{t+1:s} \in \lp{t+1,s} ~:~ \rho_{t,s}(X_{t+1:s}) \leq 0\big\} \,.
\end{equation}
and we use the notation $A^{\rho}_t$ for the acceptance set of the one-step risk measure $\rho_t$.

Next, we \cha{define a robustification of strong \tc\, dynamic risk measures that are normalised, monotone, and translation invariant. For this we first recall the notion of strong time-consistency and refer to \Cref{sec:tc} for a detailed discussion of notions of time-consistencies.

\begin{definition}[Strong time-consistency]
The dynamic risk measure $\{\rho_{t,T} \}_{t\in\mT}$ is strong \tc\, if for all $\tT$ and $X_{t+1:T}\in\lp{t+1,T}$ it holds 
\begin{equation*}
    \rho_{t,T} (X_{t+1:T}) = \rho_{t,T} \big( X_{t+1:s}+ \rho_{s,T}(X_{s+1:T}) \big)\,
    \quad \forall \;\; s \in \{t+1, \ldots, T-1\}\,.
\end{equation*}
\end{definition}

It is well-known that strong \tc\, dynamic risk measures that are normalised, monotone, and translation invariant admit a recursive representation as one-step risk measures; recalled next. 

\begin{theorem}[Recursive Relation -- \cite{cheridito2006dynamic, ruszczynski2010risk}]
    \label{thm:recursion}
    Let $\{\rho_{t,T}\}_{\tT}$ be a normalised, monotone, and translation invariant dynamic risk measure. 
    Then $\{\rho_{t,T}\}_{\tT}$ is strong \tc\footnote{\cha{This theorem also holds if strong \tc\, is replaced by order \tc\, (see \Cref{def-time-consistency-rm} for details), since these notions of time-consistencies are equivalent when the dynamic risk measure is normalised, monotone, and translation invariant  (see also \Cref{fig:time-consistencies}}).} if and only if there exists a family of one-step risk measures $\{\rho_{t}\}_{\tT}$ that are normalised, monotone, and translation invariant, such that for all $\tT$ and all $X_{t+1:T}\in\lp{t+1,T}$
    \begin{equation}
    	\rho_{t,T} (X_{t+1:T}) = 
    	\rho_{t} \Big( X_{t+1} +
    	\rho_{t+1} \big( X_{t+2} +
    	\cdots +
    	\rho_{T-1} ( X_{T} ) \cdots \big) \Big)\,. \label{eq:dynamic-risk}
    \end{equation}
\end{theorem}
As the proposed robustification is defined as} the largest (worst-case) value the dynamic risk measure can attain when evaluated at random variables in an uncertainty set\cha{, we require} the following standing assumption. 

\begin{assumption}
 All considered dynamic uncertainty sets are proper.  
\end{assumption}

\cha{
By \Cref{thm:recursion}, a normalised, monotone, translation invariant, and strong t.c. dynamic risk measure can be represented by a family of one-step risk measures that are normalised, monotone, and translation invariant. Thus, we propose to robustify at each time point the corresponding one-step risk measure by taking the worst-case within an uncertainty set.
}

\begin{definition}[Dynamic Robust Risk Measure]\label{defin R}

Let $\u$ be a dynamic uncertainty set and $\tp$ a family of normalised, monotone, and translation invariant one-step risk measures. Then we define a \textbf{dynamic robust risk measure} $R^{\u,\rho}:= \{R^{\u,\rho}_{t,T}\}_{t \in \mT}$ on $\mT$ by a sequence of conditional robust risk measures $\left\{R^{\u,\rho}_{t,T}\right\}_{\tT}$, where for each $t< s$, $\cha{\,t\in \mT ,s \in \mbT}$, the conditional robust risk measure $R^{\u,\rho}_{t,s}$ is a mapping 
$R^{\u,\rho}_{t,s}\colon  L^\infty_{t+1:s} \to {L^\infty_t}$ 
 given by
\begin{align*}
     R_{t,s}^{\u,\rho}\left(X_{t+1:s}\right)
     :=
      \esssup \, \Big\{  \rho_t (Y) \in L^\infty_t ~: ~Y \in u_{t+1} \big(X_{t+1: s}\big)  \Big\}\,.
\end{align*}
\end{definition}

Note that by definition, i.e. by taking the essential supremum, each conditional robust risk measure $R_{t,s}^{\u, \rho}$ is $\mF_t$-measurable. Moreover, any conditional robust risk measure belongs to the class of conditional risk measures and we say that a conditional robust risk measure satisfies a property in \cref{properties-rm}, if $R^{\u,\rho}_{t,s}$ satisfies it. Analogous to dynamic risk measures, we call the conditional robust risk measures $R^{\u,\rho}_{t,t+1}\colon  L^\infty_{t+1} \to {L^\infty_t}$, $t \in \mT$, a one-step (conditional) robust risk measure, and we denote it simply by $R^{\u,\rho}_t$, i.e. $R^{\u,\rho}_t (\cdot):= R^{\u,\rho}_{t, t+1}(\cdot)$.    
We drop the superscripts when there is no confusion on the dynamic uncertainty set or the family of one-step risk measures considered. That is, we write $R_{t,s}(\cdot):= R_{t,s}^{\u,\rho}(\cdot)$ and $R: = R^{\u, \rho}$.

\begin{lemma}
    Any dynamic robust risk measure is finite.
\end{lemma}
\proof{Proof.} 
As $u_{t+1}(X_{t+1:T})$ is bounded,  there exist a constant $c\in\R$ such that for all $Y \in u_{t+1}(X_{t+1:T})$, it holds that $Y \leq c$. By monotonicity of  $\rho_t$, we obtain that $R_{t,T}^{\u,\rho}\left(X_{t+1:s}\right)     =
      \esssup  \big\{  \rho_t (Y) \in L^\infty_t ~: ~Y \in u_{t+1} \big(X_{t+1: s}\big)  \big\} \leq \rho_t(c)<\infty$. 
\hfill \Halmos
\endproof
\vspace{1em}

\cha{A simple example is to set the time-$t$ uncertainty set to be the identity, i.e.}
$u_{t} (X_{t:T}) = \{ X_{t}\}$, $\tbT$. \cha{In this case there is no robustification, as} the conditional robust risk measure reduces to the conditional risk measure, that is, $R^{\u, \rho}_t (X_{t+1:T}) = \rho_t(X_{t+1})$.

\cha{Next, we provide representations of the acceptance sets of dynamic robust risk measures, illustrating that the dynamic risk measure can be interpreted as a capital requirement. }

\begin{proposition}[Acceptance Sets]
The acceptance set of a conditional robust risk measure $R_{t,T}^{\u, \rho}$, $\tT$, has representation
\begin{align*}
     A_{t,T}^R 
     & 
     =
    \Big\{ X_{t+1:T} \in L^\infty_{t+1,T} ~:~ \rho_t \left(Y\right) \leq 0, \quad \forall \; Y \in u_{t+1} \left(X_{t+1:T}\right) \Big\} 
    \\
    &=
    \Big\{ X_{t+1:T} \in L^\infty_{t+1,T} ~:~ u_{t+1} \left(X_{t+1:T}\right) \subseteq A^{\rho}_t\Big\}\,.
\end{align*}
\cha{If $R$ is translation invariant, it follows that
\begin{equation}\label{prop:eq:capital-requirement}
    R_{t,T} (X_{t+1:T} ) = \essinf \Big\{ m \in \lp{t} \; :\; X_{t+1:T}  - m \in A_{t,T}^R \Big\}\,. 
\end{equation}
\vspace*{-1.5em} 
}
\end{proposition}
\proof{Proof.} 
Since $R_{t,T}^{\u, \rho}$ is a conditional robust risk measure its acceptance set is given by \eqref{eq:def-acceptance-rm}.
The first representation of the acceptance set follows by the definition of the robust risk measure as an essential supremum over $\rho_t$, and the second by recalling the acceptance set of one-step risk measures.
\cha{The last statement follows from translation invariance of $R$.}
\hfill \Halmos
\endproof

\cha{The above proposition, in particular Equation \eqref{prop:eq:capital-requirement}, highlights that $R$ may be interpreted as a robust capital requirement, i.e., the minimum amount of capital to be added to a position in order to comply to regulated capital requirements. We observe that the time-$t$ acceptance set of $R$ consists of stochastic processes $X_{t+1:T}$, whose projection at time $t$, $X_t$, and any random variable in its uncertainty set is acceptable. Thus, in the case when the dynamic uncertainty set is static, the acceptance set of $R$ is contained in the acceptance set of the one-step risk measure, $A_{t,T}^R \subseteq \A$, indicating that fewer positions are acceptable. 
}

\cha{When robustifying a normalised, monotone, and translation invariant dynamic risk measure some of its properties may get lost. In the} next statement \cha{we investigate} which properties of the dynamic uncertainty sets induce the corresponding properties of the dynamic robust risk measure. 

\begin{proposition}[Induced Properties]\label{prop:u_to_R} 
Let $R$ be a dynamic robust risk measure with dynamic uncertainty set $\u$ and one-step risk measures $\{\rho_t\}_{\tT}$. 
Then the following holds
\begin{enumerate}[label = \roman*)]
    \item\label{u normalized - R normalize} 
    If $\u$ is normalised, then  $R$ is normalised.

    \item\label{u monotone - R monotone} 
    If $\u$ is monotone or order preserving, then  $R$ is monotone.
        
    \item\label{u trans -> R trans} 
    If $\u$ is translation invariant, then $R$ is translation invariant.
    
    \item\label{u one step - R one-step} 
    If $\u$ is static, then $R_{t,T} (\cdot) = R_{t}(\cdot)$ for all $\tT$.
    
    \item\label{u local - R local} 
    If $\u$ is local, then $R$ is local.
    
    \item  \label{u ph - R ph} 
    Let $\{\rho_t\}_{\tT}$ be positive homogeneous. If $\u$ is positive homogeneous, then $R_{t,T}(\lambda X_{t+1:T}) = \lambda R_{t,T}( X_{t+1:T}) + 1_{\lambda = 0 } R_{t,T}(0)$ for all $0 \leq \lambda \in \lp{t}$ and $X_{t+1:T}\in\lp{t+1,T}$ and $\tT$. If, moreover, $\u$ is normalised, then $R$ is  positive homogeneous.

    \item \label{u conv - R conv} Let $\{\rho_t\}_{\tT}$ be convex. If $u_t (\lambda X_{t:T} + (1-\lambda)Y_{t:T}) \subseteq \lambda u_t (X_{t:T}) + (1-\lambda) u_t(Y_{t:T})$ for all $\lambda \in \lp{t-1}$ with $ 0\leq \lambda\leq 1$ and for all $X_{t:T},\, Y_{t:T} \in \lp{t,T}$ and $\cha{\tbT}$, then $R$ is convex.

    \item \label{u suba- R sub}
    Let $\{\rho_t\}_{\tT}$ be sub-additive. If for all $Z \in u_t (X_{t:T} + Y_{t:T}) $ there exists $X' \in u_t (X_{t:T}) $ and $Y' \in u_t ( Y_{t:T})$  such that $Z \leq X' + Y'$, for all $X_{t:T},\, Y_{t:T} \in \lp{t,T}$ and $\cha{\tbT}$,
    then $R $ is sub-additive.
    
    \item \label{u sub - R sub} Let $\{\rho_t\}_{\tT}$ be sub-additive. If $u_t (X_{t:T} + Y_{t:T}) \subseteq u_t (X_{t:T}) + u_t(Y_{t:T})$ for all $X_{t:T},\, Y_{t:T} \in \lp{t,T}$ and $\cha{\tbT}$, then $R$ is sub-additive.
    
    \item \label{u conc - R conc} Let $\{\rho_t\}_{\tT}$ be concave. If $ \lambda u_t (X_{t:T}) + (1-\lambda) u_t(Y_{t:T}) \subseteq u_t (\lambda X_{t:T} + (1-\lambda)Y_{t:T})$, for all $\lambda \in \lp{t-1}$ with $0\leq \lambda\leq 1$ and $X_{t:T},\, Y_{t:T} \in \lp{t,T}$ and $\cha{\tbT}$, then $R$ is concave.
    
    \item \label{u supe- R sup}
    Let $\{\rho_t\}_{\tT}$ be super-additive. If for all $Z \in u_t (X_{t:T} + Y_{t:T}) $ there exists $X' \in u_t (X_{t:T}) $  and $Y' \in u_t ( Y_{t:T})$ such that $Z \geq X' + Y'$, for all $X_{t:T},\, Y_{t:T} \in \lp{t,T}$ and $\cha{\tbT}$, then $R $ is super-additive.
      
    \item \label{u sup - R sup} Let $\{\rho_t\}_{\tT}$ be super-additive. If $u_t (X_{t:T}) + u_t(Y_{t:T}) \subseteq u_t (X_{t:T} + Y_{t:T})$, for all $X_{t:T},\, Y_{t:T} \in \lp{t,T}$ and $\cha{\tbT}$, then $R$ is super-additive.

    \item \label{u lin - R lin } Let $\{\rho_t\}_{\tT}$ be additive. If $u_t (X_{t:T} + Y_{t:T}) = u_t (X_{t:T}) + u_t(Y_{t:T})$ for all $X_{t:T},\, Y_{t:T} \in \lp{t,T}$ and $\cha{\tbT}$, then $R$ is additive.
    
    \item \label{u star - R star} Let $\{\rho_t\}_{\tT}$ and $\u$ be star-shaped. Then $R_{t,T}(\lambda X_{t+1:T}) \leq \lambda R_{t,T}( X_{t+1:T}) + 1_{\lambda = 0 } R_{t,T}(0)$ for all $ \lambda \in \lp{t}$ with $0 \leq \lambda \leq 1$, $X_{t:T}\in \lp{t,T}$ and $\tT$. If moreover $0\in u_t(0) $ for all $\cha{\tbT}$, then $R$ is star-shaped.
\end{enumerate}
\end{proposition}

\proof{Proof.}
Throughout the proofs we simply write $R_{t,s}^{\u,\rho}\left(X_{t+1:s}\right)=\esssup \, \{  \rho_t (Y)  \;:\; Y \in u_{t+1} (X_{t+1: s})  \}$, where $\rho_t(Y) \in L^\infty_t $ is understood. Recall that by definition of $R$ the one-step conditional risk measures $\tp$ are normalised, monotone, and translation invariant, and thus also local, see e.g., Proposition 3.3 in \cite{cheridito2006dynamic}.

\cref{u normalized - R normalize}, since $\u$ is normalised it holds that $ u_{t}(0) = 0$ for all $\tbT$. Therefore
\begin{align*}
R_{t,T}(0) = \esssup \big\{\rho_t(Y): \; Y = 0 \big\} =\rho_t(0)
= 0\,, \quad \forall \; \tT\,,
\end{align*}
and $R$ is normalised.

\cref{u monotone - R monotone}, let $\tT$ and $X_{t+1:T} \leq Y_{t+1:T}$, with $X_{t+1:T}$, $Y_{t+1:T}\in \lp{t+1,T}$. If $\u$ is monotone then $u_{t+1} (X_{t+1:T}) \subseteq u_{t+1} (Y_{t+1:T})$. If $\u$ is order preserving, then for each $Z \in u_{t+1}(X_{t+1:T}) $ there exists a $Z' \in u_{t+1}(Y_{t+1:T})$ with $Z\le Z'$. Moreover, by monotonicity of $\rho_t$, we have $ \rho_t(Z)\leq \rho(Z')$. Thus, in both cases, we obtain
\begin{align*}
R_{t,T}(X_{t+1:T}) 
&=
\esssup \{\rho_t(Z) : Z \in u_{t+1}(X_{t+1:T})\}
\\
&\leq
\esssup \{\rho_t(Z) : Z \in u_{t+1}(Y_{t+1:T})\}
= R_{t,T}(Y_{t+1:T})\,,
\end{align*}
and $R_{t,T}$ is monotone.

\cref{u trans -> R trans}, let $\u$ be translation invariant. Then for $\tT$ and $Z \in \lp{t}$, we have $u_{t+1}(X_{t+1:T} + Z) = u_{t+1}(X_{t+1:T}) + Z$ and, using translation invariance of $\rho_t$, that
\begin{align*}
R_{t,T}(X_{t+1:T} + Z) 
&= \esssup \{\rho_t(Y) : Y \in u_{t+1}(X_{t+1:T} )+Z \}\\
&= \esssup \{\rho_t(Y + Z) : Y \in u_{t+1}(X_{t+1:T})\}\\
&= R_{t,T}(X_{t+1:T}) +Z\, ,
\end{align*}
and $R_{t,T}$ is translation invariant.

\cref{u one step - R one-step}, 
since $\u$ is static, we obtain for all $\tT$ that $u_{t+1}(X_{t+1:T}) = u_{t+1} ( X_{t+1})$ and 
\begin{align*}
R_{t,T}(X_{t+1:T}) = 
\esssup \{ \rho_t (Y) : Y \in u_{t+1} ( X_{t+1} ) \} 
= 
R_t(X_{t+1}) \,,
\end{align*}
and $R_{t,T}$ is static.

\cref{u local - R local}, let $\u$ be local and  for $\tT$, let $B \in \mF_t$ and $X_{t+1:T}, Y_{t+1:T} \in \lp{t+1,T}$. Then we obtain by locality of $\u$ in the second equation that  
\begin{align*}
    R_{t,T}(1_{B}\, &   X_{t+1:T} + 1_{B^\comp}   \,Y_{t+1:T}) 
    \\
    &=
    \esssup \big\{ \rho_t (Z) : Z \in u_{t+1} ( 1_{B} \,X_{t+1:T}+ 1_{B^\comp}  \,  Y_{t+1:T}) \big\} 
    \\
    &=
    \esssup \big\{ \rho_t (Z) : Z \in 1_{B} \,u_{t+1} (  X_{t+1:T}) + 1_{B^\comp} \, u_{t+1} (   Y_{t+1:T}) \big\} 
    \\
    &=
    \esssup \big\{ \rho_t (X' + Y' ) : X' \in 1_{B}\,  u_{t+1} (  X_{t+1:T}), \; Y' \in  1_{B^\comp}\, u_{t+1} (   Y_{t+1:T}) \big\} 
    \\
    &=
    \esssup \big\{ \rho_t (1_B\, X' + 1_{B^\comp}\, Y' ) : X' \in u_{t+1} (  X_{t+1:T}), \; Y' \in   u_{t+1} (   Y_{t+1:T}) \big\} 
    \\
    &=
    \esssup \big\{ 1_B \, \rho_t (X') + 1_{B^\comp} \, \rho_t ( Y' ) : X' \in u_{t+1} (  X_{t+1:T}), \; Y' \in   u_{t+1} (   Y_{t+1:T}) \big\} 
    \\
    &=
    1_{B} \,  R_{t,T}(   X_{t+1:T}) + 
     1_{B^\comp} \,  R_{t,T}(   Y_{t+1:T})\, ,
\end{align*}
where the forth equality follows since $X' \in 1_{B} u_{t+1} (  X_{t+1:T})$ implies that $X'(\omega) = 0$ for $\omega \not\in B$, and therefore $\{X' : X' \in 1_{B} u_{t+1} (  X_{t+1:T})\} = \{X'1_{B} : X' \in  u_{t+1} (  X_{t+1:T})\}$. The fifth equality holds since $\rho_t$ is  local and we conclude that $R_{t,T}$ is local.

The proofs of Items \ref{u ph - R ph} to \ref{u lin - R lin } are delegated to Appendix \ref{app:proofs}.
\hfill\Halmos
\endproof

\begin{remark} \cha{In this paper, we view} a conditional risk measure $\rho_{t,s}$ as a mapping from $\lp{t+1,s}$ to $\lp{t}$, $s<t$, while some works in the literature define conditional risk measures on \cha{the space} $\lp{t,s}$, i.e. as  $\,\tilde{\rho}_{t,s} : \lp{t,s} \to \lp{t}$. \cha{These two notations are compatible, as by translation invariance of  $\rho_{t,s}$, we can set $ \tilde{\rho}_{t,s}(X_{t:s}):= \rho_{t,s}(X_{t+1:s}) + X_t$, which is the common definition in the literature.}
Similarly for robust conditional risk measures, \cha{we can define
\begin{equation*}
    \tilde{R}_{t,s}(X_{t:s})
        :=
    R_{t,s} (X_{t+1:s}) + X_t 
         =
         \esssup \big\{  \rho_t (Y) \in L^\infty_t : Y \in u_{t+1} ( X_{t+1: s}) + X_t  \big\}\,.
\end{equation*}
Note that $\tilde{R}_{t,s}(X_{t:s})$ assesses the risk of $X_{t:s}$ at time $t$, and that in our definition we view $X_t$ as known and only account for the uncertainty of $X_{t+1:s}$. Therefore, to simplify notation we work with $R_{t,s}\colon \lp{t+1,s} \to \lp{t}$. Notice that $R_{t,s}$ satisfies one of the properties in  \cref{properties-rm} if and only if $\tilde{R}_{t,s}$ does.}
\end{remark}

\cha{The next section is devoted to generalise \Cref{prop:u_to_R} to if and only if statements. For this, we require the notion of the largest uncertainty set that gives raise to the same dynamic robust risk measures, the \Uset.}

\subsection{Consolidated Uncertainty Set.}\label{sec:Uset}

From the definition of dynamic robust risk measures, we observe that different choices of dynamic uncertainty sets may lead to the same conditional robust risk measure. Thus, we next introduce the largest uncertainty set that yields the same dynamic robust risk measure -- termed the (dynamic) \Uset. This consolidated uncertainty set will be essential for proving if and only if statements on properties of dynamic robust risk measures.

\begin{definition}[Consolidated Uncertainty Set]\label{defin U set}
Let $R$ be a dynamic robust risk measure with dynamic uncertainty set $\u$ and one-step risk measures $\{\rho_t\}_{\tT}$. Its \textbf{\Uset} $\mathfrak{U}: = \{U_t\}_{\cha{\tbT}}$ is a collection of
time-$t$  uncertainty sets $U_t$, defined for all $\cha{\tbT}$ and $X_{t:T}\in \lp{t,T}$ by
\begin{align*}
   U_{t} (X_{t:T}) 
   :=
   \bigcup \Big\{ u_t'(X_{t:T}) \subseteq \lp{t} : \quad
   &\u' = \{u_t^{\prime}\}_{\cha{\tbT}} 
  \quad   \& \quad  
   R^{\u'}_{t-1,T}(X_{t:T}) = R_{t-1,T}^{\u}(X_{t:T})     \Big\}  \,.
\end{align*} 
\end{definition}

A consolidated uncertainty set, thus, is the largest uncertainty set such that all induced robust risk measures are a.s equal. The \Uset~ has the following properties and representations.

\begin{lemma}[Representations of Consolidated Uncertainty Sets]\label{lemma u to U}
Let $R$ be a dynamic risk measure with uncertainty set $\u$ and \Uset \; $\U$. Then it holds
for all $\tT$ and $ X_{t+1:T} \in \lp{t+1,T}$ that
\begin{enumerate}[label = \roman*)]

    \item \label{ item equality bet unc sets} 
    $  U_{t+1} (X_{t+1:T}) = \left\{ Y \in \lp{t+1}: \rho_t (Y) \leq R^{\u}_{t:T}(X_{t+1:T})  \right\}$.

    \item \label{equality of the risk measure for two unc sets} 
    $R_t^\U(X_{t+1:T}) = R_t^\u (X_{t+1:T})$.
    
    \item \label{lemma u to U: 3} 
    If $\u^*: = \{u_{s}^*\}_{s \in \mbT}$ is an uncertainty set such that $R^{\u^*}_{t,T}(X_{t+1:T})  \leq R^{\U}_{t,T}(X_{t+1:T}) $, then $u^*_{t+1}(X_{t+1:T}) \subseteq U_{t+1}(X_{t+1:T})$.
    
    \item 
    \label{lemma u to U: 4}
    $ U_{t+1} (X_{t+1:T}) = \A +  R_t^\U (X_{t+1:T})$.
\end{enumerate}

\end{lemma}

\proof{Proof.} Throughout, we fix $\tT$ and $ X_{t+1:T} \in \lp{t+1,T}\,$.

\cref{ item equality bet unc sets}, let $Z \in  \{ Y \in \lp{t+1}: \rho_t (Y) \leq R^{\u}_{t:T}(X_{t+1:T})  \}$, which implies that $\rho_t (Z) \leq R^{\u}_{t:T}(X_{t+1:T})$. Next define the dynamic uncertainty set $\u^\dagger$ by $u^\dagger_{t+1} := u_{t+1} \cup \{ Z \}$ and $u^\dagger_s := u_s $ for $ s \in \mT/ \{t+1\}$. As the robust risk measure is defined through an essential supremum, it holds that $R^{\u^\dagger}_{t:T}(X_{t+1:T}) = R^{\u}_{t:T}(X_{t+1:T})$. Therefore, we conclude $Z \in U_{t+1}(X_{t+1:T})$.

Conversely, let $Y \in U_{t+1}(X_{t+1:T})$. This means, there exists a dynamic uncertainty set $\u^\dagger := \{u_t^\dagger\}_{\tbT}$  such that $Y\in u^\dagger_{t+1}(X_{t+1:T})$ and $R^{\u^\dagger}_{t:T}(X_{t+1:T}) = R^{\u}_{t:T}(X_{t+1:T})$. By definition of the robust risk measure as the essential supremum over elements in the uncertainty set, we have $\rho_t (Y) \leq R^{\u^\dagger}_{t:T}(X_{t+1:T})$  and thus $Y \in  \{ Z \in \lp{t+1}: \rho_t (Z) \leq R^{\u}_{t:T}(X_{t+1:T}) \}$.

\cref{equality of the risk measure for two unc sets}, using the representation of $U_{t+1}$ from \cref{ item equality bet unc sets}, we obtain 
\begin{align*}
R_{t:T}^\U(X_{t+1:T}) 
&= 
\esssup \big\{ \rho_t(Y) : Y \in U_{t+1}(X_{t+1:T}) \big\} 
\\
&=
\esssup \big\{ \rho_t(Y)  : \rho_t(Y) \leq R_{t:T}^\u(X_{t+1:T}) \big\}
= 
R_{t:T}^\u(X_{t+1:T})\,.
\end{align*}

\cref{lemma u to U: 3}, let $\u^*$ be a dynamic uncertainty set such that $R^{\u^*}_{t,T}(X_{t+1:T})  \leq R^{\U}_{t,T}(X_{t+1:T}) $. For $Y \in u^*_{t+1}(X_{t+1:T})$, it holds by definition of the robust risk measure that $\rho_t (Y) \leq R^{\u^*}_{t,T}(X_{t+1:T}) \leq R_{t,T}^\U(X_{t+1:T})$. Hence, by \cref{ item equality bet unc sets} $Y \in U_{t+1}(X_{t+1:T})$.

\cref{lemma u to U: 4}, recall that $\rho_t$ is translation invariant, thus by \cref{ item equality bet unc sets}
\begin{align*}
    U_{t+1} (X_{t+1:T}) 
    &=
    \left\{ Y \in \lp{t+1}: \rho_t (Y)  \leq R^{\U}_{t:T}(X_{t+1:T})  \right\}
    \\
     &=   
     \left\{ Y \in \lp{t+1}: \rho_t \left(Y -R^{\U}_{t:T}(X_{t+1:T})\right)  \leq  0 \right\}
      \\
      &=   
      \left\{ Y +R^{\U}_{t:T}(X_{t+1:T}) \in \lp{t+1}: \rho_t (Y)   \leq  0 \right\}
      \\
      &=
      \left\{ Y  \in \lp{t+1}: \rho_t (Y)   \leq  0 \right\} +R^{\U}_{t:T}(X_{t+1:T}).
       \\
       &=
       \A +R^{\U}_{t:T}(X_{t+1:T})\,,
\end{align*}
where in the last equation we used the definition of the acceptance set of $\rho_t$.
\hfill \Halmos
\endproof

\cha{With the \Uset, we can provide necessary and sufficient characterisation of the properties of dynamic robust risk measures. By \Cref{lemma u to U} \ref{equality of the risk measure for two unc sets} we have $R^\u(\cdot) = R^\U(\cdot)$, thus, whenever we say that $R$ satisfies a property we implicitly mean that both $R^\u$ and $R^\U$ satisfy it.}

\begin{theorem}
\label{theo:equiv_R_U}
Let $R$ be a dynamic robust risk measure with dynamic uncertainty set $\u$ and one-step risk measures $\{\rho_t\}_{\tT}$, and denote by $\U$ the associated \Uset. 
Then, the following holds:
\begin{enumerate}
    \item
    $R$ is normalised if and only if $U_{t+1}(0) = \A$, for all $\tT$. \label{theo:equiv_R_U_normalized}

    \item
    $R$ is monotone if and only if  $\U$ is monotone. \label{theo:equiv_R_U_monotone}
    
    \item
    $R$ is translation invariant if and only if $\U$ is translation invariant. \label{theo:equiv_R_U_trans_inv}

    \item
    $R$ is a family one-step risk measures, i.e. $R_{t,T}(\cdot) = R_t(\cdot)$ for all $\tT$, if and only if $\U$ is static. \label{theo:equiv_R_U_static}

    \item 
    $R$ is local if and only if $\U$ is local.\label{theo:equiv_R_U_locallity}
    
    \item
    \label{theo:equiv_R_U_ph} 
    Let $\{\rho_t\}_{\tT}$ be positive homogeneous. Then $R$ satisfies $ R_{t,T}(\lambda X_{t+1:T}) = \lambda  R_{t,T}( X_{t+1:T}) + 1_{\lambda=0}\, R_{t,T}( 0) $, for all $0 \le \lambda \in \lp{t-1}$, $X_{t+1:T} \in \lp{t+1,T}$ and $\tT$, if and only if $\U$ is positive homogeneous.

    \item
    Let $\{\rho_t\}_{\tT}$ be convex. Then $R$ is convex if and only if $U_t (\lambda X_{t:T} + (1-\lambda)Y_{t:T}) \subseteq \lambda U_t (X_{t:T}) + (1-\lambda) U_t(Y_{t:T})$ for all  $\lambda \in \lp{t-1} $ with $ 0\leq \lambda\leq 1$, $X_{t:T},Y_{t:T} \in \lp{t,T}$ and $\tbT$.
        \label{theo:equiv_R_U_conv}
       
    \item
    \label{theo:equiv_R_U_sub} 
    Let $\{\rho_t\}_{\tT}$ be sub-additive. Then $R$ is sub-additive if and only if $U_t (X_{t:T} + Y_{t:T}) \subseteq U_t (X_{t:T} ) + U_t (Y_{t:T})$ for all $X_{t:T},Y_{t:T} \in \lp{t,T}$ and $\tbT$.
    
   \item
   Let $\{\rho_t\}_{\tT}$ be additive. $R$ is concave if and only if $\lambda U_t (X_{t:T}) + (1-\lambda) U_t(Y_{t:T})\subseteq U_t (\lambda X_{t:T} + (1-\lambda)Y_{t:T}) $ for all  $\lambda \in \lp{t-1} $ with $ 0\leq \lambda\leq 1$, $X_{t:T} ,Y_{t:T}\in\lp{t,T}$ and $\tbT$.
   \label{theo:equiv_R_U_conc}
        
  \item 
  Let $\{\rho_t\}_{\tT}$ be additive. $R$ is super-additive if and only if $U_t (X_{t:T} ) + U_t (Y_{t:T}) 
 \subseteq U_t (X_{t:T} + Y_{t:T}) $ for all $X_{t:T} ,Y_{t:T}\in\lp{t,T}$ and $\tbT$.
  \label{theo:equiv_R_U_sup}

    \item
    Let $\{\rho_t\}_{\tT}$ be additive. $R$ is additive if and only if $U_t ( X_{t:T} + Y_{t:T}) =  U_t (X_{t:T}) +  U_t(Y_{t:T})$ for all $X_{t:T} ,Y_{t:T}\in\lp{t,T}$ and $\tbT$.
    \label{theo:equiv_R_U_lin}
    
  \item
  \label{theo:equiv_R_U_star} 
  Let $\{\rho_t\}_{\tT}$ be positive homogeneous. $R$  satisfies $ R_{t,T} ( \lambda X_{t+1:T})   \leq \lambda R_{t,T} ( X_{t+1:T}) + 1_{\lambda = 0} R_{t,T} (0) $ for all $X_{t+1:T} \in \lp{t+1,T}$, $\lambda \in\lp{t}$ with $0 \le \lambda \le 1$ and $\tT$ if and only if $\U$ is  star-shaped.
  If $\u$ is additionally normalised, then $R$ is star-shaped if and only if $\U$ is star-shaped.
\end{enumerate}
\end{theorem}

\proof{Proof.}
The ``if'' direction of Items \ref{theo:equiv_R_U_monotone}--\ref{theo:equiv_R_U_star} follow from \cref{prop:u_to_R} by taking the underlying uncertainty set to be $\U$. For \cref{theo:equiv_R_U_star}, note that if $\rho_t$ is positive homogeneous, then it is also star-shaped. For the ``if'' direction of \cref{theo:equiv_R_U_normalized}, note that if $\U_{t+1}(X_{t+1:T}) = \A$ for all $X_{t+1:T} \in \lp{t+1}$, then by \cref{lemma u to U} \ref{lemma u to U: 4} $R_{t:T}^\U (0) = 0$ and $R$ is normalised.

Next we proof the ``only if'' direction. For this, we fix $\tT$ and $ X_{0:T} \in \lp{0:T}\,$. 

\cref{theo:equiv_R_U_normalized}, let $R$ be normalised. Then by \cref{lemma u to U} \ref{lemma u to U: 4}, we have
\begin{equation*}
 U_{t+1}(0) 
 = \A +  R_t^\U (0)
 = \A\,.
\end{equation*}

\cref{theo:equiv_R_U_monotone}, let $R$ be monotone, $Y_{t+1:T} \in \lp{t+1,T}$ with $X_{t+1:T} \leq Y_{t+1:T}$, and $Z \in U_{t+1}(X_{t+1:T})$. Then, $\rho_t(Z) \leq R^\U_{t,T} (X_{t+1:T})\leq R^\U_{t,T} (Y_{t+1:T})$, where the first inequality is due to $Z \in U_{t+1}(X_{t+1:T})$ and the second by monotonicity of $R$. Therefore $Z \in  U_{t+1}(Y_{t+1:T})$ and thus $U_{t+1}$ is monotone.

\cref{theo:equiv_R_U_trans_inv}, let $R$ be translation invariant and $Z \in \lp{t}$. Then it holds by \cref{lemma u to U} \ref{ item equality bet unc sets} that 
\begin{align*}
    U_{t+1}(X_{t+1:T} + Z) 
    &=
    \left\{ Y \in \lp{t+1} : \rho_t(Y) \leq R^{\U}_{t,T}(X_{t+1:T} + Z)  \right\}
    \\ 
    &=
    \left\{ Y \in \lp{t+1} : \rho_t(Y) \leq R^{\U}_{t,T}(X_{t+1:T})+Z  \right\}  
     \\
     &=
     \left\{ Y \in \lp{t+1} : \rho_t(Y-Z) \leq R^{\U}_{t,T}(X_{t+1:T})  \right\}  
      \\
      &=
      \left\{ Y+Z \in \lp{t+1} : \rho_t(Y) \leq R^{\U}_{t,T}(X_{t+1:T})  \right\}
    \\
    &= 
    U_{t+1}(X_{t:T}) + Z\,,
\end{align*}
where the second and third by  translation invariance of $R$ and $\rho_t$, respectively, and last equalities from \cref{lemma u to U} \ref{ item equality bet unc sets}. Thus, $U_{t+1}$ is translation invariant.

\cref{theo:equiv_R_U_static}, let $R_{t,T}(\cdot) = R_t(\cdot)$ for all $\tT$, then $R^{\U}_{t,T}(X_{t+1:T})   = R^{\U}_{t,T}(X_{t+1})  $ and
 \begin{align*}
    U_{t+1}(X_{t+1:T}) 
    &=
    \left\{ Y \in \lp{t+1} : \rho_t(Y) \leq R^{\U}_{t}(X_{t+1})  \right\} =
    U_{t+1}(X_{t+1})\,,
\end{align*}
and $U_{t+1}$ is static.

\cref{theo:equiv_R_U_locallity}, let $R$ be local, $B \in \mF_t$, and $Y_{t+1:T}\in\lp{t+1,T}$. Then, using \cref{lemma u to U} \ref{lemma u to U: 4} in the first and last equation and locality of $R$ in the second, we have that
\begin{align*}
     U_{t+1} (1_B\,  X_{t+1:T} + 1_{B^\comp}\, Y_{t+1:T}) 
     &= 
     \A +  R^{\U}_{t,T} (1_B\,  X_{t+1:T} + 1_{B^\comp}\, Y_{t+1:T})
     \\
     &=
     \A +  1_B \, R^{\U}_{t,T} (X_{t+1:T}) + 1_{B^\comp} \, R^{\U}_{t,T}(Y_{t+1:T})
     \\
     &=
     1_B\,  \left(\A +   R^{\U}_{t,T} (X_{t+1:T}) \right)+ 1_{B^\comp}\,  \left(\A + R^{\U}_{t,T}(Y_{t+1:T})\right)
     \\
     &=
     1_B\,  U_{t+1} ( X_{t+1:T}) + 1_{B^\comp}  \, U_{t+1} (Y_{t+1:T})\,,
\end{align*}
and $U_{t+1}$ is local.

The proofs of Items \ref{theo:equiv_R_U_ph} to \ref{theo:equiv_R_U_star} are delegated to Appendix \ref{app:proofs}.
\hfill \Halmos
\endproof


\cha{While the above theorem characterises the properties of $R$ via its \Uset~$\U$, a dynamic robust risk measure is typically defined through a dynamic uncertainty set (e.g., the ones considered in \Cref{sec:examples}), and not its consolidated one. Thus, we next collect how properties of $\u$ translate to properties of $\U$}.
\begin{corollary}\label{cor:u_to_U}
Let $\u$ be a dynamic uncertainty set with \Uset ~ $\U$. Then,  
\begin{enumerate}

    \item If $\u$ satisfies one of the \cref{property-u} \ref{property:proper}, \ref{property:order}--\ref{property:pos-hom}, then $\U$ satisfies it. 
    
    \item\label{prop:u_to_U_normalized}  If $\u$ is normalised, then $U_{t+1}(0) = \A $ for all $\tT$.
    
    \item\label{prop:u_to_U_monotonicity} If $\u$ respects order preservation or monotonicity, then $\U$ is monotone and order preserving.

     \item \label{prop:u_to_U_suba} Let the $\{\rho_t\}_{\tT}$ be sub-additive. If $Z \in u_t (X_{t:T} + Y_{t:T}) $ implies that there is $X' \in u_t (X_{t:T}) $  and $Y' \in u_t ( Y_{t:T})$  such that $Z \leq X' + Y'$, then $U_t (X_{t:T} + Y_{t:T}) \subseteq U_t (X_{t:T} ) + U_t (Y_{t:T})  $.

       \item \label{prop:u_to_U_supe} Let $\{\rho_t\}_{\tT}$ be additive. If  $Z \geq X' + Y'$ for  any $X' \in u_t (X_{t:T}) $  and $Y' \in u_t ( Y_{t:T})$, implies that  $Z \in u_t (X_{t:T} + Y_{t:T}) $, then $U_t (X_{t:T} + Y_{t:T}) \supseteq U_t (X_{t:T} ) + U_t (Y_{t:T})  $.
       
     \end{enumerate}

\end{corollary}

\proof{Proof.}
\cref{property-u} \ref{property:proper}, let $\u$ be proper, since $\u \subseteq \U$, we have that $\U$ is non-empty. Moreover, $\u$ is bounded from above, which implies that $R$ is also bounded, and thus $\U$ needs to be bounded from above. For \cref{prop:u_to_U_monotonicity}, note that monotonicity implies the property of order preserving.
The other statements follow from the fact that by \cref{prop:u_to_R}, a property of $\u$ implies the corresponding property in $R$, which by \cref{theo:equiv_R_U} implies the corresponding property in $\U$.
\hfill \Halmos
\endproof

From the above corollary, we observe that if a dynamic uncertainty set is order preserving then its \Uset~is monotone. Since $U_{t+1}$ contains all $\mF_t$-measurable random variables $Y$ with smaller risk (\cref{lemma u to U}, \cref{ item equality bet unc sets}), monotonicity is indeed a desirable property of $\U$; while order preservation is suitable for $\u$.

Unfortunately, for a well-behaved dynamic uncertainty set $\u$, one can construct a dynamic uncertainty set $\u^*$ such that $\U = \U^*$, but $\u^*$ does not satisfy the \cha{same properties as $\u$}.

\begin{proposition}\label{prop:equiv_set_const}
Let $\u$ be an uncertainty set and $\U$ the associated \Uset. Then there exist an uncertainty set $\u^*$ such that $\U = \U^*$ and
\begin{enumerate}[label=$\roman*)$]
    \item \label{prop:equiv_set_const_normalized} $\u$ is normalised and $\u^*$ is not normalised.
    
    \item \label{prop:equiv_set_const_monotone} $\u$ is order preserving and $\u^*$ is not order preserving.
    
    \item \label{prop:equiv_set_const_translation} $\u$ is translation invariant and $\u^*$ is not translation invariant.

\end{enumerate}
\end{proposition}

\proof{Proof.}
    \cref{prop:equiv_set_const_normalized}, take $\u^* = \U$.

\cref{prop:equiv_set_const_monotone}, let $X_{t+1:T} \leq Y_{t+1:T}$ and $W\in\lp{t+1}$ be such that $\rho_t (W) = R_{t:T}(Y_{t+1:T}) $. Define $u_{t+1}^* (Y_{t+1:T}) :=\{W\}$ and $u_{t+1}^*(X_{t+1:T}) := \{ Z \in \lp{t+1} : \rho(Z) = R_{t:T}(X_{t+1:T}) \text{ and } Z\not\leq W \} $ and $u_s^* := u_s$ for all $s \neq t+1$.

\cref{prop:equiv_set_const_translation}, fix $Y_{t:T}\in \lp{t,T}$ and define for all $\tbT$
\[ u_t^* (X_{t:T}) := 
\begin{cases}
    u_t(X_{t:T}) \quad & X_{t:T}\leq Y_{t:T},
    \\ U_t(X_{t:T}) &  \text{otherwise}.
\end{cases} \]
Clearly $\u^*$ is not translation invariant. 
\hfill \Halmos
\endproof

\section{Time-consistent Dynamic Robust Risk Measures.}\label{sec:tc}

In the dynamic setting, notions of time-consistency are of utmost importance when, e.g., optimising dynamic risk measures, see e.g., \cite{acciaio2011dynamic,bielecki2017survey} for a review. The first key result in this section is \cref{theo:equiv_R_U_time}, which provides \cha{necessary and sufficient criteria for} different notions of time-consistencies. The second is \cref{thm:recusion-DRRM}, which states that a \cha{dynamic robust risk measure is strong t.c. or \nn\,} if and only if it can be constructed \cha{recursively} via a static uncertainty set.

\subsection{{\color{black}{Notions of Time-consistencies.}}}\label{sec: notion of tc}
\cha{This section is devoted to when time-consistency is preserved by robustification. For this, we} first define notions of time-consistencies for dynamic uncertainty sets which we then relate to time-consistencies of dynamic robust risk measures. \cha{Several researchers proposed different definitions of time-consistency and it is not our intention to provide an exhaustive review of this concept, as we put a focus on time-consistencies that result in a recursive representation. We provide  interpretation of time-consistencies discussed in this section after their definition and refer the reader referred to \cite{bielecki2017survey}, which offers an encompassing survey.
}

\begin{definition}[Time-Consistency of Dynamic Uncertainty Sets]\label{def-time-consistency-u}
Let $R$ be a dynamic robust risk measure with dynamic uncertainty set $\u$ and one-step risk measures $\{\rho_t\}_{\tT}$. Then, $\u$ is
\begin{enumerate}[label = \roman*)]
    \item\label{item-strong-time}
    \textbf{Strong t.c.}, if for all $\cha{\tbT}$ and $X_{t:T} \in \lp{t,T}$ it holds
\begin{equation*}
    u_t (X_{t:T}) = u_t \big( X_{t:s}+ R_{s,T}(X_{s+1:T}) \big)\,,
    \quad \forall \;\; s \in \{t, \ldots, T-1\}\,.
\end{equation*}

\item\label{item-order-time}
\textbf{Order t.c.}, if for all $\cha{\tbT}$ and $X_{t:T}, Y_{t:T} \in \lp{t,T}$ that satisfy 
\begin{equation*}
   X_{t:s} = Y_{t:s} \quad \text{ and } \quad u_{s+1}\left(X_{s+1:T}\right)\, \,\subseteq\, \,u_{s+1}\left(Y_{s+1:T}\right)
 \end{equation*}
for some $s\in\{t, \ldots, T-1\}$, it holds 
 \begin{equation*}
     u_{t}\left(X_{t:T}\right) \subseteq u_{t}\left(Y_{t:T}\right) \,.
 \end{equation*}

\item\label{item-rejection-time}
\textbf{Rejection t.c.}, if for all $\cha{\tbT}$ and $X_{t:T}\in\lp{t,T}$ with $X_t \ge 0\,$, 
\begin{equation*}
 0 \in u_{t+1}\left(X_{t+1:T}\right) 
\quad \text{implies} \quad 
     0 \in u_{t}\left(X_{t:T}\right) \,.
 \end{equation*}

\item\label{item-non-norm-time}
\textbf{\cha{Weak recursive}}, if for all $\cha{\tbT}$ and $X_{t:T}\in\lp{t,T}$ it holds 
\begin{equation*}
    u_{t}\left(X_{t:T}\right) = u_t \big( X_{t:s} + R_{s,T}\left(X_{s+1:T}\right) -  R_{s,T}(0) \big)\,,  
    \quad \forall \;\; s \in \{t, \ldots , T-1 \}\, .
\end{equation*}

\item\label{item-weak-time}
\textbf{Weak t.c.}, if for all $\cha{\tbT}$ and $X_{t:T}\in\lp{t,T}$ it holds 
\begin{equation*}
         u_t \big( X_{t:s}+ R_{s,T}\left(X_{s+1:T}\right) \big)\, \subseteq\,  u_t \left(X_{t:T}\right)\,,
     \quad \forall \;\; s \in \{t, \ldots , T-1 \}\, .
\end{equation*}

\item\label{item-prudence-time} \textbf{Prudent}, if for all $\cha{\tbT/\{T\}}$ and $X_{t:T}\in\lp{t,T}$ it holds 
\begin{equation*}
  X_t + R_{t,T}\left(X_{t+1:T}\right)- R_{t,T}(0)\, \in\, u_t\left(X_{t:T}\right)\,,
 \end{equation*} 
 \cha{and $X_T \in u_T(X_T)$.}
\end{enumerate}
\end{definition}

\cha{
Note that prudence implies $X_t \in u_t (X_{t})$. Furthermore, in \cref{prop:u_R_time}, we show that prudence results in the dynamic robust risk measure always dominating the dynamic risk measure. }

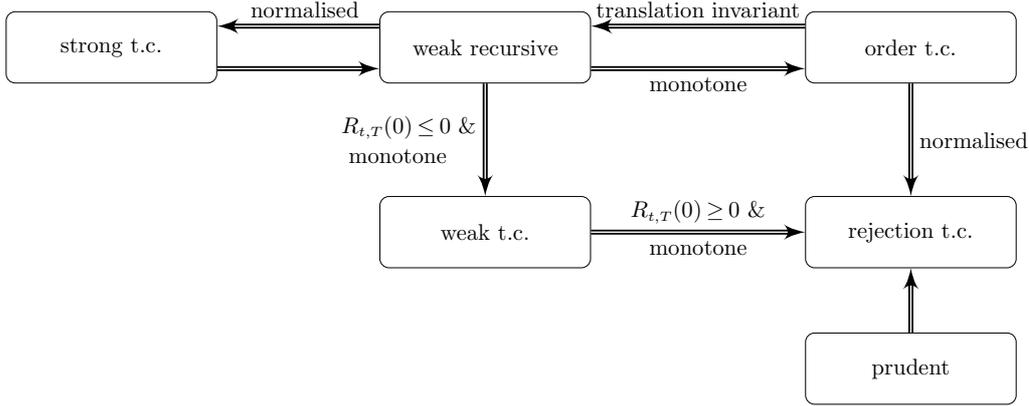
\begin{figure}
   
    \begin{center}

  \vspace*{1em}
  \resizebox{0.95\textwidth}{!}{
\begin{tikzpicture}[node distance = 1.5cm, auto,>=stealth]

 \tikzstyle{quadri} = [rectangle, draw, fill=none, text width=3cm, text centered,text=black , rounded corners, minimum height=3em]
 \tikzstyle{line} = [draw, thick, color=black, -latex']

 \node[quadri] (S) {\small{strong \tc}};
 
\node[quadri, right=2.5cm of S ] (N) {\small{\nn}};

\node[quadri,below=1.75cm  of N] (W) {\small{weak \tc}};

\node[quadri, right=3.3cm of N ] (O) {\small{order \tc}};
 
\node[quadri,below = 1.75cm  of O] (R) {\small{rejection \tc}};

\node[quadri, below =1cm  of R] (P) {\small{prudent}};

\path[line,double] ([yshift=-0.33cm]S.east) --  ([yshift=-0.33cm]N.west);
\path[line,double] ([yshift=0.33cm]N.west) -- node [sloped, above, pos=0.5] {\small{\;\;normalised}} ([yshift=0.33cm]S.east);

\path[line,double] ([yshift=0.33cm]O.west) --node [above, above] {\small{translation invariant}}  ([yshift=0.33cm]N.east);
 \path[line,double] ([yshift=-0.33cm]N.east) -- node [ below] {\small{monotone}} ([yshift=-0.33cm]O.west);

\path[line,double] (N) --node [left, align=left] { \small{$ R_{t,T} (0) \leq 0$ \&} \\ \small{ monotone}} (W);

\path[line,double] (O) --node [right] {\small{normalised}} (R);

\path[line,double] (P) --(R);

\path[line,double] (W) --node [above] {\small{$ R_{t,T} (0) \geq 0$} \& } node[below]{\small{monotone}} (R);

 \end{tikzpicture}
 }
     \caption{Time-consistencies for a \Uset~$\U$ and for a dynamic robust risk measure $R$. By \cref{theo:equiv_R_U_time}, the diagram holds for both $\U$ and $R$, with the exception of prudence, which is only defined for $\U$.}\label{fig:time-consistencies}
 \end{center}

\end{figure}

Next, we recall different notions of time-consistencies of dynamic risk measures and define a new version called \cha{weak recursiveness. We also refer to \Cref{fig:time-consistencies} on how these different time-consistencies are connected. }

\begin{definition}[Time-Consistency of Dynamic Risk Measures]\label{def-time-consistency-rm}
Let $R$ be a dynamic robust risk measure with dynamic uncertainty set $\u$ and one-step risk measures $\{\rho_t\}_{\tT}$. Then, $R$ is
\begin{enumerate}[label = \roman*)]

\item\textbf{Order t.c.}, if for all $\tT$ and $X_{t+1:T}, Y_{t+1:T}\in\lp{t+1,T}$ that satisfy
\begin{equation*}
   X_{t+1:s} = Y_{t+1:s} \quad \text{ and } \quad R_{s,T}(X_{s+1:T}) \leq R_{s,T}(Y_{s+1:T})
      \end{equation*}
for some $s\in\{t+1, \ldots, T-1 \}$, it holds 
 \begin{equation*}
     R_{t,T}(X_{t+1:T}) \leq R_{t,T}(Y_{t+1:T}) \,.
 \end{equation*}

\item \textbf{Rejection t.c.}, if for all $\tT$ and $X_{t+1:T}\in\lp{t+1,T}$ with $X_{t+1} \geq 0$ 
\begin{equation*}
R_{t+1,T}(X_{t+2:T}) \geq 0 \quad \text{implies} \quad 
        R_{t,T}(X_{t+1:T}) \geq 0\,.
 \end{equation*}
 
\item \textbf{\cha{Weak recursive}}, if for all $\tT$ and $X_{t+1:T}\in\lp{t+1,T}$ it holds
\begin{equation*}
     R_{t,T}(X_{t+1:T}) =   R_{t,T} \big( X_{t+1:s} + R_{s,T}(X_{s+1:T}) -  R_{s,T}(0) \big) 
     \quad \forall \; s \in \{t+1, \ldots, T-1\}\, .
\end{equation*}

\item \textbf{Weak t.c.}, if for all $\tT$ and $X_{t+1:T}\in\lp{t+1,T}$ it holds
\begin{equation*}
        R_{t,T} \big( X_{t+1:s} + R_{s,T}(X_{s+1:T}) \big) \leq R_{t,T}(X_{t+1:T}) \,.
\end{equation*}
\end{enumerate}
\end{definition}

While the literature on time-consistency is extensive, different names and definitions are used and ``time-consistency'' is often referred to without further distinction, e.g., \cite{bion2020fully,cheridito2006dynamic}. In many works, the concept of strong \tc\, refers to a property where the risk measure exhibits a recursive nature, also called ``recursivity''  \cite{bielecki2017survey,detlefsen2005conditional,epstein2010ambiguity}. Order \tc\, denotes congruent preferences of decisions with respect to ordering or ranking across different time periods. \cha{In the context of preferences relations, order \tc\, appears in the works of \cite{epstein2003recursive,Maccheroni2006JEF,wang2003conditional}}. Rejection \tc\, means that an unacceptable outcome tomorrow is also unacceptable today. In \cite{tutsch2008update}, this property is referred to as ``rejection \tc\, with respect to ${0}$", which under translation invariance is equivalent to weak rejection \tc\, considered in \cite{bielecki2017survey}. 
While the definition of weak recursiveness is new, a related concept is considered in \cite{bion2020fully, nunno2023fully}, who study fully dynamic risk measures.
The property of prudence means that dynamic uncertainty sets contain the future risk they assesses. 

\cha{In the literature,} conditional risk measures are often  assumed to be normalised, monotone, and translation invariant, in which case strong t.c., order t.c.,  and \nn\,  are equivalent and often referred to simply as ``time-consistent" \cite{detlefsen2005conditional,frittelli2006risk,riedel2004dynamic,roorda2005coherent,delbaen2006structure}. \cha{However, when robustifying dynamic risk measures these properties may get lost, making it necessary to study them separately. We next relate the new notion of \nn\,to strong and order t.c. For this we make to following observation.

\begin{lemma}\label{lemma:normalise-tc}
Let $R$ be \nn\, and define the normalised version of $R$ by $\tilde{R}_{t,T} (\cdot) := {R_{t,T}} (\cdot) - {R_{t,T}} (0) $ for $\tT$. Then $\tilde{R}$ is strong t.c. 
\end{lemma}
As discussed in \Cref{sec:examples}, many examples of uncertainty sets yield dynamic robust risk measures that are not normalised, e.g. induced by the Wasserstein distance  \cite{luo2020distributionally,gao2022distributionally,esfahani2015data,luo2019decomposition}.
In the context of fully dynamic risk measures, the work \cite{nunno2023fully} analysis how concepts of time-consistency, h-longevity, and
restriction are connected to  normalisation. We provide an example of a dynamic robust risk measure that is strong \tc\, but not normalised, in \Cref{ex:NN-strong-TC-RM}.

The following alternative characterisation of weak recursiveness illustrates its difference to order time-consistency. Indeed, as seen in \Cref{fig:time-consistencies}, neither of these notion imply the other.}
\begin{lemma}\label{lemma:on weak time-consisten}
      $R$ is \nn\,if and only if for all $\tT$ and
      $X_{t+1:T},Y_{t+1:T}\in\lp{t+1,T}$ that satisfy
      \begin{equation*}
      X_{t+1:s} + R_{s,T}(X_{s+1:T}) = Y_{t+1:s} +R_{s,T}(Y_{s+1:T})    
      \end{equation*}
      for all  $s\in\{t+1,\ldots, T-1\}$, it holds that 
      \begin{equation*}
          R_{t,T}(X_{t+1:T}) = R_{t,T}(Y_{t+1:T}) \,.
      \end{equation*} 
   \end{lemma}
\proof{Proof.}
For the  if direction let $X_{t+1:T}\in\lp{t+1,T}$ and define $Y_{t+1:T}:= X_{t+1:s} + R_{s,T}(X_{s+1:T}) -R_{s,T}(0) $ which satisfies, since $Y_{s+1:T} = 0$, that $R_{s,T}(Y_{s+1:T}) = R_{s,T}(0)$. Therefore $X_{t+1:T} \,, Y_{t+1:T}$ satisfy that $Y_{t+1:s} +R_{s,T}(Y_{s+1:T})  = X_{t+1:s} + R_{s,T}(X_{s+1:T}) -R_{s,T}(0)  +R_{s,T}(0)  = X_{t+1:s} + R_{s,T}(X_{s+1:T})  $, and thus the assumption in the statement. Therefore $R_{t,T}(X_{t+1:T}) = R_{t,T}(Y_{t+1:T}) = R_{t,T}( X_{t+1:s} + R_{s,T}(X_{s+1:T}) -R_{s,T}(0) )$ and $R$ is \nn.

For the  ``only if'' direction assume that $R$ is \nn\,and $ X_{t+1:T}\,, Y_{t+1:T}\in\lp{t+1,T}$ satisfy $X_{t+1:s} + R_{s,T}(X_{s+1:T}) = Y_{t+1:s} +R_{s,T}(Y_{s+1:T})  $. Then we have, for any $s\in\{t+1, \ldots, T-1\}$ that 
    \begin{align*}
      R_{t,T}(X_{t+1:T})  &=   R_{t,T}(X_{t+1:s} + R_{s,T}(X_{s+1:T}) -  R_{s,T}(0))
      \\ &= R_{t,T}(Y_{t+1:s} + R_{s,T}(Y_{s+1:T}) -  R_{s,T}(0)) 
      =R_{t,T}(Y_{t+1:T})\,,
    \end{align*}
which concludes the proof.
\hfill \Halmos
\endproof

With these notions of time-consistencies \cha{at hand}, we now state \cha{the properties} of $\u$ \cha{that yield \tc\, dynamic robust risk measures}. 
\begin{proposition}[Induced Time-Consistencies]\label{prop:u_R_time}
Let $R$ be a dynamic robust risk measure with dynamic uncertainty set $\u$. Then, the following holds:
\begin{enumerate}
    
    \item\label{prop:u_R_time_strong}  
    If $\u $ is strong \tc, then $R$ is strong t.c.
    
    \item\label{prop:u_R_time_order}  
    Let $\u $ be order t.c. Let $X_{t+1:T}, Y_{t+1,T} \in \lp{t+1,T}$ satisfy $X_{t+1:s} = Y_{t+1:s} $ and $u_{s+1}(X_{s+1:T}) \subseteq u_{s+1}(Y_{s+1:T})$, then it holds that $  R_{t,T}(X_{t+1:T}) \leq R_{t,T}(Y_{t+1:T})$.

    \item\label{prop:u_R_time_non norm} 
    If $\u$ is \nn, then $R$ is \nn.
    
    \item\label{prop:u_R_time_wea} 
    If $\u$ is weak \tc, then $R$ is weak t.c.
    
    \item\label{prop:u_R_prudency} 
    If $\u$ is prudent, then $R_{t,T}(X_{t+1:T}) \ge \rho_t\big(X_{t+1}+R_{t+1,T}(X_{t+2:T})\big)$ for all $X_{t+1:T}\in\lp{t+1,T}$. In particular, it holds that
    \begin{equation*}
    R_{0,T} (X_{1:T})
    \geq 
    \rho_0 \big(X_1 + R_{1,T} (X_{2:T})\big) 
    \;\geq  \dots \geq\;
    \rho_{0} \circ \dots \circ \rho_{T-1}\left(\sum_{s 
 =1}^T X_s\right)\,.    
    \end{equation*}
\end{enumerate}
\end{proposition}

\proof{Proof.} Throughout, we let $\tT$ and $X_{0:T}\in \lp{0:T}$.\\
\cref{prop:u_R_time_strong}, let $\u$ be strong \tc, then we have for all $s \in \{t+1, \ldots, T-1\}$ that $u_{t+1} ( X_{t+1:T}) = u_{t+1} ( X_{t+1:s}+ R_{s,T}(X_{s+1:T}) )$. Since the two  uncertainty sets are the same, they give rise to the same robust risk measures, i.e. $R_{t,T}(X_{t+1:T}) =R_{t,T}\big(X_{t+1:s} + R_{s,T}(X_{s+1:T})\big)$, and $R$ is strong t.c.

\cref{prop:u_R_time_order}, let $\u$ be order time-consistency and $X_{t+1:T}, Y_{t+1:T}$ such that $u_{t+1}(X_{t+1:T}) \subseteq u_{t+1}(Y_{t+1:T})$. Then, by definition of $R$, it holds that $  R_{t,T}(X_{t+1:T}) \leq R_{t,T}(Y_{t+1:T})$.

\cref{prop:u_R_time_non norm}, let $\u$ be \nn, then $u_{t+1} ( X_{t+1:T}) = u_{t+1} \big( X_{t+1:s}+ R_{s,T}(X_{s+1:T}) - R_{s,T}(0 )\big)$, which implies that the corresponding robust risk measures are equal, i.e., that $R_{t,T}(X_{t+1:T})  =R_{t,T}\left(X_{t+1:s} + R_{s,T}\big(X_{s+1:T}\big) - R_{s,T}(0 )\right)$, and $R$ is order t.c.

\cref{prop:u_R_time_wea}, let $\u$ be weak \tc, then $u_{t+1} \big( X_{t+1:s}+ R_{s,T}(X_{s+1:T})\big) 
 \subseteq u_{t+1} ( X_{t+1:T}) $ implies the ordering of the robust risk measures, i.e. $   R_{t,T}(X_{t+1:T}) \ge R_{t,T}\big(X_{t+1:s} + R_{s,T}(X_{s+1:T} )\big)$, and $R$ is weak t.c.

\cref{prop:u_R_prudency}, let $\u$ be prudent, then $X_{t+1} + R_{t+1:T}(X_{t+2:T}) \in u_{t+1}(X_{t+1:T})$ and 
\begin{equation*}
    R_{t:T}(X_{t+1:T})
    = 
    \esssup\{\rho_t(Y) : Y \in u_{t+1}(X_{t+1:T})\}
    \ge
    \rho_t\big(X_{t+1} + R_{t+1:T}(X_{t+2:T})\big)\,.
\end{equation*}
Applying the above inequality recursively concludes the proof.
\hfill \Halmos
\endproof

Next, we provide one of the key results, uniquely connecting time-consistencies of \Uset{s}~with those of the dynamic robust risk measure. 

\begin{theorem}
\label{theo:equiv_R_U_time}
Let $R$ be a dynamic robust risk measure with dynamic uncertainty set $\u$ and \Uset \; $\U$. Then, the following holds:
\begin{enumerate}
    \item\label{theo:equiv_R_U_time_strong} 
    $R $ is strong \tc\, if and only if $\U$ is strong t.c.
 
    \item\label{theo:equiv_R_U_time_order}  $R$ is order \tc\, if and only if $\U$ is order t.c.

    \item\label{theo:equiv_R_U_time_rejection} $R$ is rejection \tc\, if and only if  $\U$ is rejection t.c.
    
     \item\label{theo:equiv_R_U_time_non norm} $R$ is \nn\, if and only if $\U$ is \nn.
     
    \item\label{theo:equiv_R_U_time_wea}  $R$ is weak \tc\, if and only if $\U$ is weak t.c.
    \item\label{theo:equiv_R_U_time_prudent}  $R$ satisfies $R_{t,T}(X_{t+1:T}) \ge \rho_t\big(X_{t+1}+R_{t+1,T}(X_{t+2:T})\big)$, for all $\tT$ and $X_{t+1:T} \in \lp{t+1,T}$, if and only if $\U$ is prudent.
\end{enumerate}

\end{theorem}

\proof{Proof.}
 The ``if'' direction of 
Items \ref{theo:equiv_R_U_time_strong}, \ref{theo:equiv_R_U_time_non norm}, \ref{theo:equiv_R_U_time_wea}, and \ref{theo:equiv_R_U_time_prudent} follow from \cref{prop:u_R_time} with $\U=\u$. Throughout, we let $\tT$ and $X_{0:T} \in \lp{0,T}$.

\cref{theo:equiv_R_U_time_strong}, let $R$ be strong \tc, then
    \begin{align*}
        U_{t+1}(X_{t+1:T}) &= \left\{  Y \in \lp{t+1}: \rho_t(Y) \leq R_{t,T}(X_{t+1:T})  \right\}
        \\ 
        &=
        \left\{  Y \in \lp{t+1}: \rho_t(Y) \leq R_{t,T}\big( X_{t+1:s}+ R_{s,T}(X_{s+1:T})\big)  \right\}
        \\ &= U_{t+1}\big( X_{t+1:s}+ R_{s,T}(X_{s+1:T})\big)\,,
    \end{align*}
    and $\U$ is strong t.c.

\cref{theo:equiv_R_U_time_order}, first let $\U$ be order \tc\, and $X_{t+1:s} = Y_{t+1:s} $ with $R^{\U}_{s,T}(X_{s+1:T}) \leq R^{\U}_{s,T}(Y_{s+1:T})$. Then $U_{s+1} (X_{s+1:T}) = \{ Z \in \lp{s+1} : \rho_{s}(Z) \leq  R^{\U}_{s,T}(X_{s+1:T}) \} \subseteq \{ Z \in \lp{s+1} : \rho_{s}(Z) \leq  R^{\U}_{s,T}(Y_{s+1:T}) \} = U_{s+1} (Y_{s+1:T}) $. Hence, by order time-consistency of $\U$, we have $U_{t+1} (X_{t+1:T}) \subseteq U_{t+1} (Y_{t+1:T}) $, which implies that $R^{\U}_{t,T}(X_{t+1:T})  \le R^{\U}_{t,T}(Y_{t+1:T})$.

Second, let $R$ be order \tc\, and for $s \in\{t+1, \ldots, T-1\}$, let $X_{t+1:s} = Y_{t+1:s} $ with $ U_{s+1}(X_{s+1:T}) \subseteq U_{s+1}(Y_{s+1:T})$. Since $R$ is defined through a supremum, we have $ R_{s,T}(X_{s+1:T}) \leq R_{s,T}(Y_{s+1:T})$ and, by order time-consistency of $R$, $ R_{t,T}(X_{t+1:T}) \leq R_{t,T}(Y_{t+1:T})$. Hence,
 \begin{align*}
        U_{t+1}(X_{t+1:T}) &= \left\{  Y \in \lp{t+1}: \rho_t(Y) \leq R_{t,T}(X_{t+1:T})  \right\}
        \\ &\subseteq \left\{  Y \in \lp{t+1}: \rho_t(Y) \leq R_{t,T}(Y_{t+1:T})  \right\}
        = U_{t+1}( Y_{t+1:T})\,,
    \end{align*}
and $\U$ is order t.c.

\cref{theo:equiv_R_U_time_rejection}, note that for any $\tT$, we have $0 \in U_{t+1} (X_{t+1:T})$ if and only if $R_{t,T}(X_{t+1:T}) \geq 0$. First, let $\U$ be rejection \tc\, and let $X_{t+1:T}\in\lp{t+1,T}$ with $X_{t+1} \ge 0$ and $R_{t+1,T}(X_{t+2:T}) \geq 0$. Then, $0 \in  U_{t+2} (X_{t+2:T})$  and by rejection time-consistency of $\U$, $0 \in U_{t+1}(X_{t+1:T})$. Hence, $R_{t,T}(X_{t+1:T}) \geq 0$ and $R$ is rejection t.c.

Second, let $R$ be rejection \tc\, and let $X_{t:T}\in\lp{t,T}$ with $X_t \ge 0$ and $0 \in U_{t+1}(X_{t+1:T})$. Then, $R_{t,T}(X_{t+1:T}) \geq 0$  and by rejection time-consistency of $R$, $R_{t-1,T}(X_{t:T}) \geq 0$. Hence, $0 \in U_{t}(X_{t:T})$ and $\U$ is rejection t.c.

\cref{theo:equiv_R_U_time_non norm}, let $R$ be \nn. Therefore all $s \in \{t+1, \ldots, T-1\}$,
    \begin{align*}
        U_{t+1}(X_{t+1:T}) &= \left\{  Y \in \lp{t+1}: \rho_t(Y) \leq R_{t,T}(X_{t+1:T})  \right\}
        \\ &= \left\{  Y \in \lp{t+1}: \rho_t(Y) \leq R_{t,T}\big( X_{t+1:s}+ R_{s,T}(X_{s+1:T}) -R_{s,T}(0) \big)  \right\}
        \\
        &=
        U_{t+1}\big( X_{t+1:s}+ R_{s,T}(X_{s+1:T}) -R_{s,T}(0) \big)\,,
    \end{align*}
and $\U$ is \nn.

\cref{theo:equiv_R_U_time_wea}, let $R$ be weak \tc, then for all $s \in \{t+1, \ldots, T-1\}$,
 \begin{align*}
        U_{t+1}(X_{t+1:T}) 
        &=
        \left\{  Y \in \lp{t+1}: \rho_t(Y) \leq R_{t,T}(X_{t+1:T})  \right\}
        \\ 
        &\supseteq
        \left\{  Y \in \lp{t+1}: \rho_t(Y) \leq R_{t,T}\big( X_{t+1:s}+ R_{s,T}(X_{s+1:T})\big)  \right\}
        \\ &= U_{t+1}\big( X_{t+1:s}+ R_{s,T}(X_{s+1:T})\big)\,,
    \end{align*}
    and $\U$ is weak t.c.

\cref{theo:equiv_R_U_time_prudent}, let $R$ satisfy $R_{t,T}(X_{t+1:T}) \ge \rho_t\big(X_{t+1}+R_{t+1,T}(X_{t+2:T})\big)$. Then, we have that $X_{t+1} + R_{t+1,T}(X_{t+2:T})  \in \left\{  Y \in \lp{t+1}: \rho_t(Y) \leq R_{t,T}(X_{t+1:T})  \right\} = U_{t+1}(X_{t+1:T})$ and $\U$ is prudent.
\hfill \Halmos
\endproof

The next corollary collects which notions of time-consistencies of $\U$ are implied by those of $\u$. 
\begin{corollary}\label{cor:u-U-time}
Let $\u$ be a dynamic uncertainty set with \Uset ~ $\U$. If $\u$ satisfies one of the time-consistencies in \cref{def-time-consistency-u} \ref{item-strong-time}, \ref{item-non-norm-time}, 
\ref{item-weak-time}, \ref{item-prudence-time},  then $\U$ satisfies it. 
\end{corollary}

\proof{Proof.}
 By \cref{prop:u_R_time}, each property in $\u$ implies the analogous property in $R$, which by \cref{theo:equiv_R_U_time} implies the corresponding property in $\U$.
 \hfill \Halmos
 \endproof

We conclude this section by providing connections and implications of the different notions of time-consistencies, which are illustrated in \cref{fig:time-consistencies}. Note that by \cref{theo:equiv_R_U_time} most of the statements also hold for dynamic robust risk measures.

\begin{proposition}[Time-consistencies of Consolidated Uncertainty Sets]\label{prop:consistency_static}
Let $R$ be a dynamic robust risk measure with dynamic uncertainty set $\u$ and \Uset \; $\U$. 
Then it holds  that:
     \begin{enumerate}
        \item\label{prop:consistency_strong imply non norm}
        If $\U$ is strong \tc, then it is \nn.
     
         \item\label{prop:consistency_equiv non norm and strong} 
         Let $\U$ satisfy $U_{t+1} (0) = \A$ for all $\tT\}$. If $\U$ is \nn, then it is strong time-consistent.

         \item\label{prop:consistency_stong implies indiference in R(0)} 
          \cha{ If $\U$ is strong t.c., then  $ R_{t,T} \big( X_{t+1:T} + \lambda R_{s,T} (0) \big) =   R_{t,T} ( X_{t+1:T} ) $, for all $X_{t+1:T} \in \lp{t+1,T}$ and $ \lambda \in \mathbb{Z} $.}
          
         \item\label{prop:consistency_non norm to order time} 
         If $\U$ is monotone and \nn, then it is order t.c.

         \item\label{prop:consistency_order to non norm} 
         If $\U$ is translation invariant and order \tc, then it is \nn.

        \item\label{prop:consistency_non norm to weak} 
         Let $R_{t,T}(0) \leq 0$ for all $\tT$. If $\U$ is monotone and \nn, then it is weak t.c.
         
         \item\label{prop:consistency_weak to rejection} Let $\U$ be monotone and $0 \in U_{t}(0)$ for all $\tbT$. If $\U$ is weak \tc, then it is rejection t.c.
         \item\label{prop:consistency_prudence to rejection}
         If $\U$ is prudent, then it is rejection t.c.
         \item\label{prop:consistency_order to rejection} 
         Let $U_{t+1}(0) = \A$ for all $\tT$. If $\U$ is order \tc, then it is rejection t.c.
     \end{enumerate}
     
\end{proposition}

\proof{Proof.}
\cref{prop:consistency_strong imply non norm},
let $\U$ be strong \tc, then, by \cref{theo:equiv_R_U_time} \cref{theo:equiv_R_U_time_strong}, $R$ is strong t.c. Next, let $ X_{t+1:T}\,, Y_{t+1:T}\in\lp{t+1,T}$ satisfy $X_{t+1:s} + R_{s,T}(X_{s+1:T}) = Y_{t+1:s} +R_{s,T}(Y_{s+1:T})  $ for all $s\in\{t+1, \ldots, T-1\}$. Then by strong time-consistency of $R$,
\begin{align*}
      R_{t,T}(X_{t+1:T})  =   R_{t,T}\big(X_{t+1:s} + R_{s,T}(X_{s+1:T}) \big)
     = R_{t,T}(Y_{t+1:s} + R_{s,T}(Y_{s+1:T})) 
      =R_{t,T}(Y_{t+1:T})\,.
    \end{align*}
Thus, by \cref{lemma:on weak time-consisten}, $R$ is \nn, and applying  \cref{theo:equiv_R_U_time} \cref{theo:equiv_R_U_time_non norm} yields the claim.

\cref{prop:consistency_equiv non norm and strong}, by \cref{theo:equiv_R_U}, $U_{t+1}(0) = \A$ is equivalent to $R$ being normalised. Moreover, if $R$ is normalised, then strong time-consistency  and \nn\, are equivalent.

\cha{
\cref{prop:consistency_stong implies indiference in R(0)}, by \cref{prop:consistency_strong imply non norm} $R$ is \nn. Next, fix $X_{t+1:T}\in \lp{t+1,T}$ and define $Y_{t+1:T} := X_{t+1:T} - R_{s,T}(0)$.
Then, we obtain by first applying weak recursiveness, then the definition of $Y$, and strong t.c., 
\begin{subequations}\label{pf:eq-nn-strong}
\begin{align}
      R_{t,T}(X_{t+1:T})  &=   R_{t,T}\big(X_{t+1:s} + R_{s,T}(X_{s+1:T})  - R_{s,T} (0) \big) 
       \\
     &= R_{t,T}(Y_{t+1:s} + R_{s,T}(Y_{s+1:T})) 
      \\ 
      & =R_{t,T}(Y_{t+1:T})
      \\ 
      &=  R_{t,T}\big(X_{t+1:T} - R_{s,T} (0)\big)\,.
    \end{align}
    \end{subequations}
    Iteratively applying Equations \eqref{pf:eq-nn-strong} to  $\tilde{X}_{t+1:T}: = X_{t+1:T} - R_{s,T} (0)$ yields that  $R_{t,T}(X_{t+1:T}) =  R_{t,T}(X_{t+1:T} - \lambda R_{s,T} (0)) $, for any positive integer $\lambda$. 
    
    Next, we apply Equations \eqref{pf:eq-nn-strong} to $\tilde{X}_{t+1:T}: = X_{t+1:T} + R_{s,T} (0)$ which gives
    \begin{equation}\label{pf:eq-nn-strong2}
    R_{t,T}\big(X_{t+1:T} + R_{s,T}(0)\big)=R_{t,T}\big(X_{t+1:T} + R_{s,T}(0)- R_{s,T}(0)\big)  = R_{t,T}\big(X_{t+1:T}\big)\,.
    \end{equation}
Iteratively applying Equation \eqref{pf:eq-nn-strong2} to $\tilde{X}_{t+1:T}: = X_{t+1:T} + R_{s,T} (0)$ yields that  $R_{t,T}(X_{t+1:T}) =  R_{t,T}(X_{t+1:T} + \lambda R_{s,T} (0)) $, for any positive integer $\lambda$. Combining these results concludes the proof.
}

\cref{prop:consistency_non norm to order time}, for $t\le s$, let $X_{t:s} = Y_{t:s}$ such that $U_{s+1} (X_{s+1:T}) \subseteq U_{s+1} (Y_{s+1:T})$. This implies that $R_{s,T}(X_{s+1:T}) \leq R_{s,T}(Y_{s+1:T})$ and therefore 
\begin{equation*}
  X_{t:s} + R_{s,T}(X_{s+1:T}) - R_{s,T}(0) \;\leq \;Y_{t:s} + R_{s,T}(Y_{s+1:T}) - R_{s,T}(0)\,.  
\end{equation*}
By monotonicity and weak recursiveness of $\U$, it holds
\begin{align*}
U_{t}(X_{t:T}) 
    &= U_t \big(  X_{t:s} + R_{s,T}(X_{s+1:T}) - R_{s,T}(0) \big) 
    \\
    &\subseteq 
    U_t \big(   Y_{t:s} + R_{s,T}(Y_{s+1:T}) - R_{s,T}(0)) = U_t (Y_{t:T} \big)\,,      
\end{align*}
which implies that $\U$ is order t.c.

\cref{prop:consistency_order to non norm},  by \cref{theo:equiv_R_U,theo:equiv_R_U_time}, $R$ is translation invariant and order t.c. We show that $R$ is \nn. For $s \in \{t+1, \ldots T-1\}$ define $Y_{t+1:T}$ by $Y_{t+1:s}:= X_{t+1:s}$, $Y_{s+1}:= R_{s,T}(X_{s+1:T}) - R_{s,T}(0)$, and $Y_{s+2:T}:=0$. Then, by translation invariance of $R$ and noting that $R$ is not necessarily normalised, we obtain
\begin{align*}
    R_{s,T}(Y_{s+1:T}) 
    &=
    R_{s,T}\big(R_{s,T}(X_{s+1:T}) -R_{s,T}(0) \big) 
    \\
    &= 
    R_{s,T}(0) + R_{s,T}(X_{s+1:T})  -R_{s,T}(0) = R_{s,T}(X_{s+1:T})\,.
\end{align*}
Hence, $Y_{t+1:s} = X_{t+1:s}$ and $R_{s,T}(X_{s+1:T}) = R_{s,T}(Y_{s+1:T})$, thus by order time-consistency of $R$ it holds that 
$$R_{t,T}(X_{t+1:T}) =R_{t,T}(Y_{t+1:T})  = R_{t,T}\big(X_{t+1:s} + R_{s,T}(X_{s+1:T}) - R_{s,T}(0)\big)\,,$$
and $R$ is \nn. \cref{theo:equiv_R_U_time} concludes the proof.

 \cref{prop:consistency_non norm to weak}, by \nn\, of $\U$, we have for all $s \in \{t, \ldots, T-1\}$ that $U_t(X_{t:T}) =U_t(X_{t:s} + R_{s,T}(X_{s+1:T}) - R_{s,T}(0)) \supseteq U_t(X_{t:s} + R_{s,T}(X_{s+1:T}))   $, where the set inclusion is due to monotonicity of $\U$ and since $R_{s,T}(0) \le 0$. Thus, we conclude that $\U$ is weak t.c.

\cref{prop:consistency_weak to rejection}, note that by \cref{theo:equiv_R_U,theo:equiv_R_U_time}, $R$ is monotone and weak t.c. Let $0 \in U_{t+2}(X_{t+2:T}) $, and $X_{t+1} \geq 0$. This implies, by normalisation of $\rho_{t+1}$, that
\begin{equation}\label{pf:eq:R-non-neg}
     R_{t+1,T}(X_{t+2:T}) 
     =
     \sup\Big\{\rho_{t+1}(Y) : Y \in U_{t+2}(X_{t+2:T})\Big\}
     \ge 
     \rho_{t+1}(0) = 0\,.
\end{equation}
Next, by weak time-consistency (with $s = t+1$), then monotonicity of $R$, and then applying \eqref{pf:eq:R-non-neg}, we have
\begin{align*}
R_{t,T}(X_{t+1:T}) 
    &\geq 
    R_{t,T} \big( X_{t+1} + R_{t+1,T}(X_{t+2:T}) \big)
    \\
    &\geq 
    R_{t,T} \big( R_{t+1,T}(X_{t+2:T}) \big)
    \\
    &
    \geq 
    R_{t,T} ( 0) 
    \geq 0    \,,
\end{align*}
where the last inequality follows since 
 by assumption $0 \in U_s(0)$, for all $s \in \mbT$. Thus, we conclude that $\U$ is rejection t.c.

\cref{prop:consistency_prudence to rejection}, first note that prudence implies $R_{t}(X_{t+1:T})   \geq    \rho_t\big(X_{t+1} + R_{t+1}(X_{t+2:T})\big) $.
Next, let $R_{t+1,T}(X_{t+2:T}) \geq 0  \text{ and }  X_{t+1} \geq 0$, then, monotonicity and normalisation of $\rho_t$ implies that  $R_{t}(X_{t+1:T}) \geq \rho_t(X_{t+1} + R_{t+1}(X_{t+2:T})) \geq \rho_t (0) =0$. We conclude that $R$ and hence $\U$ is rejection t.c.

\cref{prop:consistency_order to rejection}, let $0 \in U_{t+1}(X_{t+1:T}) $ and $X_t \geq 0$. Recall that by \cref{theo:equiv_R_U}, $U_{t+1}(0) = A_{\rho_{t+1}}$ implies  that $R$ is normalised. Therefore, $ R_{t,T}(0) = 0 \leq  R_{t,T}(X_{t+1:T})$,  where the inequality follows since $0 \in  U_{t+1}(X_{t+1:T}) $.
Hence $ U_{t+1}(0) \subseteq  U_{t+1}(X_{t+1:T})$. Next, define $X_{t:T}':= (0, X_{t+1:T})$ and $Y_{t:T}:=0$. Then clearly $Y_t = 0 = X_t$ and $ U_{t+1}(Y_{t+1:T}) \subseteq  U_{t+1}(X_{t+1:T})$, thus, by order time-consistency of $\U$, we have that $ U_{t}(Y_{t:T}) = U_{t}(0)\subseteq  U_{t}(X_{t:T})$. By assumption, we have 
$U_{t}(0) = \A \ni 0$, where 0 is an element of $\A$ since $\rho_t$ is normalised. We conclude that $\U$ is rejection t.c.
\hfill \Halmos
\endproof

\subsection{Construction of \tc\, Dynamic Robust Risk Measures.}\label{sub-sec:construction-time}

In this section we investigate the intrinsic connection between static uncertainty sets and \tc\, dynamic uncertainty sets. To facilitate notation, we denote by $\u^\varsigma$ a dynamic uncertainty set that possesses the property of being static. Recall that by Corollary \ref{cor:u_to_U} the corresponding \Uset~$\U^{\vs}$ is also static. 

The first result is negative, in that static uncertainty sets cannot give rise to (most of the notions of) \tc\, dynamic robust risk measures. However, one of the key results of this section (\cref{thm:recusion-DRRM}) is that dynamic robust risk measures are  \nn\, if and only if they admits a recursive representation of one-step robust conditional risk measures that arise from a static uncertainty set\cha{; generalising \Cref{thm:recursion} to the robust setting. To establish this, we require the following result on the connection of static and \tc\, dynamic uncertainty sets.}

\begin{proposition}\label{prop:static-\tc}
    Let $R$ be a dynamic robust risk measure with dynamic uncertainty set $\u$ and \Uset \; $\U$. 
Then the following hold:
     \begin{enumerate}
         \item\label{prop:consistency_static_no} 
        Let $R_{t,T}$ be surjective  for all $\tT$. If $\u^\vs$ is static, then $\U^\vs$ and $R^{\U^\vs}$ do not satisfy  the time-consistency notions of \cref{item-strong-time,item-non-norm-time,item-weak-time}   of \cref{def-time-consistency-u,def-time-consistency-rm} and \cref{item-prudence-time} of \cref{def-time-consistency-u}.
         
         \item\label{prop:consistency_static_non norm}
         $R^\U$ or equivalently $\U$ is \nn\, if and only if there exists a static \Uset \; $\U^\vs: = \{U_t^\vs\}_{\tbT}$ satisfying recursively backwards in time 
         \begin{subequations}
         \begin{align}\label{eq:U-nn-iff-static}
            \cha{U_T(X_T)} &\cha{= U_T^\vs(X_T)        
             \,, \quad \text{and}}
             \\
             U_t \left(X_{t:T}\right)
             &=
             U^\vs_t \big(X_t + R^{\U}_{t,T} \left(X_{t+1:T}\right)  -R^{\U}_{t,T} (0) \big) 
              \,,
             \quad \forall \;\;\tbT\setminus\{T\}  \,.
         \end{align}
            \end{subequations}
            
    \item\label{prop:consistency_static_strong} If $R^\U$ or equivalently $\U$ is strong \tc, then there exists a static \Uset \; $\U^\vs: = \{U_t^\vs\}_{\tbT}$ satisfying recursively backwards in time 
         \begin{align*}
            \cha{U_T(X_T)} &\cha{= U_T^\vs(X_T)        
             \,, \quad \text{and}}
             \\
             U_t \left(X_{t:T}\right)
             &=
             U^\vs_t \big(X_t + R^{\U}_{t,T} \left(X_{t+1:T}\right)  \big) 
            \,,
             \quad \forall \;\;\tbT\setminus\{T\}  \,.
         \end{align*}
\end{enumerate}
\end{proposition}

\proof{Proof.}
\cref{prop:consistency_static_no}, let $\u^\vs$ be static and strong \tc\, and fix $X_t \in \lp{t}$. Then it follows that $u_t^\vs (X_{t}) = u_t^\vs (X_{t:T}) = u_t^\vs \big(X_{t}  + R_{t,T} (X_{t+1:T})   \big) $ where $X_{t+1:T}$ is an arbitrary process in $\lp{t+1,T}$. Moreover, surjectivity of $R_{t,T}$ implies that for any $Y_t \in \lp{t}$ there exists an $X_{t+1:T}\in\lp{t+1,T}$ such that $Y_t = X_{t}  + R_{t,T} (X_{t+1:T}) $. Thus, $u_t^\vs (X_t) = u_t^\vs(Y_t)$ for any $X_t,Y_t \in \lp{t}$. This implies that $R_{t-1,T} (X_{t:T}) = C \in \lp{t-1}$ for all  $X_{t:T} \in \lp{t,T}$, which, by surjectivity of $R_{t,T}$, cannot be true. The same reasoning holds if $\u$ is \nn.

Let $\u^\vs$ be static and weak t.c. Then $u_t^\vs (X_{t}) = u_t^\vs (X_{t:T}) \supseteq u_t^\vs \big(X_{t}  + R_{t,T} (X_{t+1:T})   \big) =u_t^\vs (Y_{t})  $ for all $X_t, Y_t \in \lp{t} $. Reversing the role of $X_t$ and $Y_t$ it follows that $u_t^\vs (X_t) = u_t^\vs(Y_t)$ for all $X_t, Y_t \in \lp{t}$. This implies that $R$ is a constant contradicting surjectivity.

Finally, let $\u^\vs$ be static and prudent. By surjectivity of $R$, for any  $Y_t \in \lp{t}$ there exists a $X_{t+1:T}\in\lp{t+1,T}$ such that $Y_t =  R_{t,T} (X_{t+1:T})$. Then, by prudence of $\u^\vs$, we have $X_t + Y_t = X_t + R_{t,T} (X_{t+1:T})  \in u_t^\vs (X_{t:T})= u_t^\vs (X_t)$ for all $X_t, Y_t \in \lp{t}$. Thus, we conclude $u_t^\vs (X_t) = \lp{t}$ for all $X_t \in \lp{t}$, which implies that $\u$ cannot be proper.

\cref{prop:consistency_static_non norm}, let $\U$ be \nn\, and define for all $X_{t:T} \in \lp{t,T}$ the dynamic uncertainty set $\U^\vs: = \{U_t^\vs\}_{\tbT}$ via
\begin{align*}
    U^\vs_t (X_t + R^{\U}_{t,T} (X_{t+1:T})  -R^{\U}_{t,T} (0) )  :
     = U_t (X_{t:T})  
     =U_t \big(X_t + R^{\U}_{t,T} (X_{t+1:T})  -R^{\U}_{t,T} (0) \big)\,,
\end{align*}
where the second equality follows since $\U$ is \nn.
Furthermore, $\U^\vs$ is static since for any $Y_t \in \lp {t}$, there exists a $X_{t:T} \in \lp{t,T}$ such that $Y_t = X_t + R^{\u}_{t,T} (X_{t+1:T})  -R^{\u}_{t,T} (0)  $. Note that we do not need $R$ to be surjective, as we can choose $X_t \in \lp{t}$ arbitrarily.

Conversely, let $\U$ be given and let $\U^\vs$ be a static uncertainty satisfying \eqref{eq:U-nn-iff-static} recursively backwards in time. 
Define $Y_{t:T}$ by $Y_t:= X_{t} + R^{\U}_{t,T} \left(X_{t+1:T}\right)  -R^{\U}_{t,T} (0)$ and $Y_{t+1:T}:= 0$. Then, applying equation \eqref{eq:U-nn-iff-static} first to $Y_{t:T}$ and in the last equation to $X_{t:T}$, we obtain
\begin{align*}
    U_t\big(X_{t} + R^{\U}_{t,T} \left(X_{t+1:T}\right)  -R^{\U}_{t,T} (0) \big)
    &=U_t(Y_{t:T})
    \\
    &= 
    U^\vs_t \big(Y_t + R^{\U}_{t,T} \left(0\right)  -R^{\U}_{t,T} (0) \big)
    \\
    &=
     U^\vs_t \big(X_{t} + R^{\U}_{t,T} \left(X_{t+1:T}\right)  -R^{\U}_{t,T} (0) \big)
      \\&=    U_t \big(X_{t:T} \big)\,.
\end{align*}
Thus, $\U$ is \nn.

\cref{prop:consistency_static_strong} follows using similar reasoning as \cref{prop:consistency_static_non norm}. 
\hfill \Halmos
\endproof

\cref{prop:static-\tc} states that the \Uset~of any strong t.c. or  \nn\, 
 dynamic robust risk measure can be represented via static uncertainty sets. This provides a way to construct \tc\, dynamic robust risk measures from static uncertainty sets. Moreover, the resulting dynamic risk measure can be seen as recursively applying a robust conditional risk measure arising from a static uncertainty set.

\begin{theorem}[Construction of Dynamic Uncertainty Sets]\label{theo: static to robust dynamic}
Let $\u^\vs$ be a dynamic uncertainty set that is static and define a dynamic uncertainty set $\u:=\{u_t\}_{\tbT}$ recursively backwards in time
\begin{align*}
    \cha{u_T(X_T):}&\cha{= u_T^\vs(X_T)\,, \quad \text{and}}
    \\
    u_t(X_{t:T}) :&= u^\vs_t\big(X_t + R^{\u}_{t,T} (X_{t+1:T}) -R^{\u}_{t,T} (0) \big) \,,
    \quad \forall\;\;\tbT\setminus\{T\}  \quad \text{and}\quad X_{t:T} \in \lp{t,T}\,.
\end{align*}
Then, it holds that $ R^{\u}_{t,T}(X_{t+1:T}) = R^{\u^\vs}_{t}\big( X_{t+1} + R^{\u}_{t+1,T}(X_{t+2:T}) -R^{\u}_{t+1,T} (0) \big) $ and
\begin{equation*}
    R^{\u}_{t,T}\left(X_{t+1:T}\right) = R^{\u^\vs}_{t}\left( Y_{t+1} + 
                                        R^{\u^\vs}_{t+1}\big( Y_{t+2} +
                                            R^{\u^\vs}_{t+2}( Y_{t+3} + \ldots + R^{\u^\vs}_{T-1} (Y_T) \ldots )\big)\right)\,,
\end{equation*}
where $Y_{t} = X_{t} -R^{\u^\vs}_{t}(0)$ for all $\tT$.

Denote by $\U$ the \Uset~of $\u$, then 
\begin{enumerate}[label=$\roman*)$]
    \item\label{thm:construction-NN}
    $R^{\u}$ and $\U$ are \nn.
    \item\label{thm:construction-strong}
    If  $\u$ is normalised, then $R^{\u}$ and $\U$ are strong t.c.
    
    \item\label{thm:construction-normalised}
    $\u$ is normalised if and only if $\u^\vs$ is normalised.
    
    \item\label{thm:construction-trans}
    $\u$ is translation invariance if and only  $\u^\vs$ is translation invariant.
    
    \item\label{thm:construction-order}
    $\u$ is order preserving if and only if $\u^\vs$ is order preserving.
\end{enumerate}

\end{theorem}

\proof{Proof.}
Let $\U^\vs$ be the \Uset~ of $\u^\vs$. Then by \cref{cor:u_to_U}, $\U^\vs$ is static and $\U$ satisfies  $U_t(X_{t:T}) = U^\vs_t\big(X_t + R^{\u}_{t,T} (X_{t+1:T})-R^\u_{t,T}(0)\big)$ for all $\tbT\setminus\{T\} $. The property $ R^{\u}_{t,T}(X_{t+1:T}) = R^{\u^\vs}_{t}\big( X_{t+1} -R^{\u}_{t+1,T} (0)+ R^{\u}_{t+1,T}(X_{t+2:T})  \big)$ follows by definition of the dynamic robust risk measure and noting that $\U^\vs$ is static. The recursion follows by applying this property recursively and since $u_t(0) = u^\vs_t(0)$ for all $\tbT$.

\cref{thm:construction-NN}, \nn\, of $R^\u$ and $\U$ follows from \cref{prop:static-\tc}. \cref{thm:construction-strong}, if $\u$ is normalised, then by \cref{cor:u_to_U} $\U$ satisfies $U_{t+1}(0) = \A$ for all $\tT$. Applying \cref{prop:consistency_static} gives that $\U$ and thus also $R^\u$ are strong time-consistency.

\cref{thm:construction-normalised}, if $\u^\vs$ is normalised, then $\u$ is by definition normalised. Next, assume that $\u$ is normalised, then
\begin{equation*}
    0 = u_t(0) = u^\vs_t\big(0 + R^{\u}_{t,T} (0) -R^{\u}_{t,T} (0) \big)
    =u^\vs_t(0)\,,
\end{equation*}
and $\u^\vs$ is normalised.

\cref{thm:construction-trans}, let $\u^\vs$ be translation invariant, then we have for any $Z \in \lp{s}, \;s<t$, that
\begin{align*}
  u_t(X_{t:T} )  + Z
  &=
  u^\vs_t\big(X_t + R^{\u}_{t,T} (X_{t+1:T}) -R^{\u}_{t,T} (0) \big) +Z   
  \\
  &=
  u^\vs_t\big(X_t + R^{\u}_{t,T} (X_{t+1:T})-R^{\u}_{t,T} (0) + Z \big) 
  =  u_t(X_{t:T} +Z)  \,,
\end{align*}
and $\u^\vs$ is translation invariant. The other direction follows using similar steps.

\cref{thm:construction-order}, let $\u^\vs$ be order preserving and $X_{t:T} \leq Y_{t:T} $. Then, $u_T(Z_T) = u^\vs_T(Z_T)$, thus $u_T$ is order preserving, and, by \cref{prop:u_to_R}, $R^{\u}_{T-1,T}$ is monotone. Thus, 
\begin{equation*}
X_{T-1} + R^{\u}_{T-1,T} (X_{t+1:T}) -R^{\u}_{T-1,T} (0)\leq Y_{T-1} + R^{\u}_{T-1,T} (Y_{t+1:T})  -R^{\u}_{T-1,T} (0)\,.    
\end{equation*}
Next, let $Z \in u_{T-1} (X_{T-1,T})=u^\vs_{T-1} \big( (X_{T-1} + R^{\u}_{T-1,T} (X_{t+1:T}) -R^{\u}_{t,T} (0) \big)$, then since $u_{T-1}^\vs$ is order preserving, there exists a $W \in u^\vs_{T-1} \big( (Y_{T-1} + R^{\u}_{T-1,T} (Y_{t+1:T}) -R^{\u}_{t,T} (0) \big) = u_{T-1}(Y_{T-1:T})$ such that $Z \le W$, and thus $u_{T-1}$ is order preserving. Applying the same reasoning recursively backwards in time yields that $\u$. 

For the ``only if'' direction, we show that if $\u^\vs$ is not order preserving then $\u$ is also not order preserving. For this, let $\u^\vs$ be not order preserving. That is for some $\tbT$ and  $X_t \leq Y_t$, there does not exists $W \in u^\vs_t(Y_{t})$ satisfying $Z\leq W$ for all $Z \in u^\vs_t(X_{t}) $. 
Now define $X_t' := X_t + R^\u_{t,T}(0)$ and $Y_t' := Y_t + R^\u_{t,T}(0)$ which satisfy $X_t' \le Y_t'$. Then 
\begin{equation*}
    u_t(X_t')
    =
    u_t\big(X_t + R^\u_{t,T}(0)\big)
    =
    u_t^\vs(X_t)
\end{equation*}
and similarly $u_t(Y_t') = u_t^\vs(Y_t')$.
Since $\u$ is not order preserving, there does not exist a $W \in u^\vs_t(Y_{t})=u_t(Y_t')$ such that $Z\leq W$, for all $Z \in u^\vs_t(X_{t}) = u_t(X_t')  $. Hence, $u_t$ is not order preserving.
\hfill \Halmos
\endproof

We further obtain that any static uncertainty set gives raise to a \tc\, dynamic robust risk measures via a recursive representation.

\begin{theorem}[Recursive Relation]\label{thm:recusion-DRRM}
\cha{Let $R$ be a dynamic robust risk measure. Then the following holds: 
\begin{enumerate}[label = $\roman*)$]
    \item \label{item:nntc-recursion}
    $R$ is \nn\, if and only if there exists a static   uncertainty set $\u^\vs := \{u_t^\vs\}_{\tbT}$ such that for all $\tT$ and $X_{t+1:T}\in\lp{t+1,T}$
    \begin{equation}\label{eq:recursion-nntc}
    R_{t,T}\left(X_{t+1:T}\right) = R^{\u^\vs}_{t}\bigg( Y_{t+1} + 
                                        R^{\u^\vs}_{t+1}\Big( Y_{t+2} +
                                            R^{\u^\vs}_{t+2}\big( Y_{t+3} + \ldots + R^{\u^\vs}_{T-1} (Y_T) \ldots \big)\Big)\bigg)\,,
\end{equation}
where $Y_t  := X_{t} -R^{\u^\vs}_{t} (0)$, $\tT$, and $Y_T:=X_T$.

    \item \label{item:strongtc-recursion}
    $R$ is normalised and strong \tc\, if and only if there exists a static and normalised uncertainty set $\u^\vs := \{u_t^\vs\}_{\tbT}$ such that Equation \eqref{eq:recursion-nntc} holds with $Y_t  := X_{t}$, $\tbT$.
\end{enumerate}
}
\end{theorem}
\proof{Proof.} The ``if'' direction \cha{of both parts} is a consequence of \cref{theo: static to robust dynamic}. For the ``only if'' direction \cha{of \ref{item:nntc-recursion}}, let $R^\u$ be \nn, then by \cref{prop:static-\tc} \cref{prop:consistency_static_non norm} there exists a static \Uset \; $\U^\vs$ that satisfies Equation \eqref{eq:U-nn-iff-static}. Since $\U^\vs$ proper, we can define $\u^\vs:= \U^\vs$, which is static and proper. Applying \cref{theo: static to robust dynamic}, we obtain the recursion.

For \cha{the ``only if'' direction of \ref{item:strongtc-recursion} let } $R^\u$ be strong \tc, then there exists by \cref{prop:static-\tc} \cref{prop:consistency_static_strong} a static \Uset~$\U^\vs: = \{U_t^\vs\}_{\tbT}$ satisfying
\begin{equation}\label{eq:proof-static-strong-tc}
    U_t(X_{t:T}) = U_t^\vs\big(X_t + R_{t,T}^\U (X_{t+1:T})\big)\,, \quad 
    \forall \tbT\setminus\{T\}  \,.
\end{equation}
Since $\U^\vs$ is proper, we can define for all $\tT$
\begin{equation*}
 \u^\vs(X_t) := \U^\vs(X_t)   \,, \quad \text{if} \quad X_t \neq 0\,,
 \quad \text{and}\quad 
 \u^\vs(0):= 0\,.
\end{equation*}
Clearly, $\u^\vs$ is normalised, proper and its \Uset~is $\U^\vs$ which satisfies \eqref{eq:proof-static-strong-tc}. \cref{theo: static to robust dynamic} provides the recursion.
\hfill\Halmos \endproof

\cha{The next results uniquely connects dynamic uncertainty sets that satisfy the recursive relation.}

\begin{proposition}
    Let $\u$ and $\u'$ be two dynamic uncertainty sets. If for all $\tT$ it holds that
    \begin{equation*}
    R^{\u}_{t,T}\left(X_{t+1:T}\right) 
    = 
    R^{\u'}_{t}\bigg( X_{t+1} + 
       R^{\u'}_{t+1}\Big( X_{t+2} +
    R^{\u'}_{t+2}\big( X_{t+3} + \ldots + R^{\u'}_{T-1} (X_T) \ldots \big)\Big)\bigg)\,,
\end{equation*}
then $U_t\left(X_{t:T}\right) = U'_t\big(X_t + R^{\u}_{t,T} \left(X_{t+1:T}\right)\big) $, $\tbT\setminus\{T\}$, \cha{and $U_t(X_T) = U_T'(X_T)$.} 
\end{proposition}

\proof{Proof.}
From the recursion, we obtain for all $\tT\setminus\{T-1\} $ that $R^{\u}_{t,T}(X_{t+1:T}) = R^{\u'}_{t}( X_{t+1} + R^{\u}_{t+1,T}(X_{t+2:T}) )$. Therefore 
\begin{align*}
  U_{t+1}(X_{t+1:T}) &= \left\{ Y \in \lp{t+1} : \rho_t (Y) \leq R^{\u}_{t,T}(X_{t+1:T}) \right\}
  \\ &= \left\{ Y \in \lp{t+1} : \rho_t (Y) \leq  R^{\u'}_{t}\big( X_{t+1} + R^{\u}_{t+1,T}(X_{t+2:T}) \big) \right\} 
  \\ &= U'_{t+1}\big(X_{t+1} + R^{\u}_{t+1,T} (X_{t+2:T})\big)\,,
\end{align*}
which concludes the proof.
\hfill \Halmos
\endproof

\section{Examples of Dynamic Robust Risk Measures.}\label{sec:example-rm}

All dynamic uncertainty sets discussed in \cref{sec:examples} can be used to define dynamic robust risk measures. By \cref{prop:u_to_R}, properties of the dynamic robust risk measure follow from those of the dynamic uncertainty set. As seen in \cref{theo: static to robust dynamic}, if the strong  time-consistency  or weak recursiveness is desirable, then static uncertainty sets are called for. From \cref{tab:u_set_properties}, we observe that $f$-divergences, and in particular the KL-divergence, may lead to dynamic robust risk measures that are strong t.c. while uncertainty sets generated by semi-norms and Wasserstein distances may result in weak recursiveness. The first example illustrates \cha{a construction of dynamic robust risk measures using the} dynamic uncertainty set of \cref{example: norms}. \cha{\cref{ex:NN-strong-TC-RM} provides a dynamic robust risk measure that is strong t.c. and not normalised. 
}
\cref{ex:dual-rep} discusses dynamic robust risk measures induced by the dual representation of convex risk measures.

\begin{example}[Semi-Norm on Random Variables]\label{example: norms two}
The dynamic uncertainty set $\u^{||\cdot||}$ in \cref{example: norms}, \cref{eq:ex:u-seminorm}  gives rise to the dynamic robust risk measure
\begin{align*}
    R_t^{\u^{||\cdot||}}(X_{t+1})
    &=
    \esssup_{\Vert X_{t+1}- Y\Vert \le \ep_{X_t}} \rho_t(Y)\,, 
    \quad \forall \;  \tT\,.
\end{align*}

If $ \Vert \ep_{X_t} \Vert \leq \ep_{X_t} $ and $\sup\{ Y :  \Vert Y \Vert \leq \ep_{X_t} \} = \ep_{X_t}  $, or $u^{||\cdot||}$ is given by \eqref{eq:ex:u-seminorm-spacial},
then by monotonicity and translation invariance of $\rho_t$ we obtain that 
\begin{align*}
    R_t^{\u^{||\cdot||}}(X_{t+1})
    &=
    \esssup \big\{\rho_t(Y +X_{t+1}) \in \lp{t} : \Vert Y    \Vert \leq \ep_{X_t} \big\}
    =
    \rho_t(X_{t+1}) + \ep_{X_t} \,,
\end{align*}
thus the robust risk measure additively decomposes into the risk of $X_{t+1}$ and its uncertainty $\ep_{X_{t}}$.
By \cref{theo: static to robust dynamic}, $\u^{||\cdot||}$ can be used to construct a \nn\, dynamic uncertainty set $\u': = \{u'_t\}_{\tbT}$ through the recursive procedure \cha{$u_T'(X_T) := u_T^{||\cdot||}(X_T)$ and} 
\begin{align*}
    u'_t(X_{t:T}) := u_t^{||\cdot||}\Big(X_t + R^{\u^{||\cdot||}}_{t} \big(X_{t+1:T}\big) -R^{\u^{||\cdot||}}_{t} (0)\Big)\,, \quad \cha{t\in\mbT/\{T\}\,.}
\end{align*}
where the resulting robust risk measure is 
\begin{multline*}
        R^{\u'}_{t,T}\big(X_{t+1:T}\big) 
        = R^{\u^{||\cdot||}}_{t}\bigg( X_{t+1} -R^{\u^{||\cdot||}}_{t+1} (0) + 
           R^{\u^{||\cdot||}}_{t+1}\Big( X_{t+2} -R^{\u^{||\cdot||}}_{t+2} (0) +
           \\
             \cdots + R^{\u^{||\cdot||}}_{T-1} (X_T) \cdots \Big)\bigg)\,.
\end{multline*}
For the trivial norm and $0 < \ep < 1$, $u^{||\cdot||}$ reduces to the identity. In this case, there is no uncertainty, thus $R^{\u^{||\cdot||}} (X_t) = \rho_t(X_t)$ and $ R^{\u'}_{t,T}(X_{t+1:T}) = \rho_t \circ \dots \circ \rho_{T-1} \left( \sum_{i=t+1}^T X_i \right) $.
Alternatively, if we use a $p$-norm, including the supremum norm, we have $R^{u^{||\cdot||}} (X_t) = \rho_t(X_t) + \ep_{X_t}$ and \begin{equation}\label{eq:ex:recusion}
R^{u'}_{t,T}(X_{t+1:T})     =
    \rho_t \circ \dots \circ \rho_{T-1} \left( \sum_{i=t+1}^T X_i \right)
    +  \rho_t \circ \dots \circ \rho_{T-1} \left( \sum_{i=t+1}^T (\ep_{X_i} - \ep_0)  \right)\,.   
\end{equation} 
The term $\sum_{i=t+1}^T (\ep_{X_i} - \ep_0) $ may represent decreasing uncertainty over time and that longer time horizons are more uncertainty. 
\end{example}

\cha{\begin{example}[Strong \tc\, but not normalised]\label{ex:NN-strong-TC-RM}
We construct a dynamic robust risk measure that is strong \tc\, but not normalised. Let the one-step risk measures be the conditional expectation, $\rho_t(\cdot):= \E[\cdot| \mF_t]$, and define the time-$t$ uncertainty set 
\begin{equation*}
    u_t(X_{t:T}) := \Big\{ Y \in \lp{t} : \, \E \Big[ Y- \sum_{i=t}^T X_i \Big| \mF_t   \Big] \leq \ep_{t-1} \Big\},
\end{equation*}
    where $\ep_{t-1} \in \lp{t-1}$ is a non-degenerate random variable. The corresponding dynamic robust risk measure is 
    \begin{equation}\label{ex:strong-tc-non-norm}
        R_{t,T}(X_{t+1:T}) = \E \Big[  \sum_{i=t+1}^T X_i \Big| \mF_t   \Big] +\ep_t \,, \qquad \tT\,,
    \end{equation}
    which is not normalised as $R_{t,T}(0) = \ep_t \neq 0$. Moreover, $R$ is \nn\, and further satisfies 
        \begin{equation*}
        R_{t,T}(X_{t+1:s} + R_{s,T}(X_{s+1:T})   ) 
         = 
        \E \Big[  \sum_{i=t+1}^T X_i \Big| \mF_t   \Big] + \E[\ep_s | \mF_t] + \ep_t \,.
    \end{equation*}
    From the above equations, we observe that $R$ is strong \tc\, if and only if $\E[\ep_t | \mF_t]=0$, for all $\tT$. 
    Next, we consider the normalised version  of \eqref{ex:strong-tc-non-norm}, that is $\tilde{R}_{t,T}(\cdot):= R_{t,T}(\cdot) - R_{t,T}(0) $. By \Cref{lemma:normalise-tc}, $\tilde{R}$ is strong \tc\, however, the robustification is lost. Indeed $\tilde{R}_{t,T}(X_{t+1:T}) = \E [  \sum_{i=t+1}^T X_i | \mF_t   ]$.   
   
\end{example}
}

\cha{
\begin{example}[Uncertainty sets induced by dual representation]\label{ex:dual-rep} 
Let $\mathcal{P}$ be the set of probability measures that are absolutely continuous w.r.t. $\P$ and $\{\rho_t\}_{\tT}$ a convex and Fatou-continuous sequence of one-step risk measures. Define by $\alpha_t^{min} : \mathcal{P} \rightarrow \lp{t}\cup\{\infty\}$ the \emph{minimal penalty functions} of $\{\rho_t\}_{\tT}$ given by
    \[\alpha^{min}_t(\Q) := \esssup \big\{ \E_\Q [ X_{t+1} | \mF_t ]: \rho_t(X_{t+1}) \leq 0 \big\}\,,  \]
where $\E_\Q$ denotes the expectation under $\Q\in \mathcal{P}$.\footnote{An example is the conditional entropic risk measure with the (scaled) conditional KL divergence as penalty function,  i.e. $\rho_t (X_{t+1}) = \frac{1}{\beta} \log \E \left[ \exp{ (\beta X_{t+1}) }| \mF_t \right]$ and $\alpha^{min}_t (\Q) = \frac{1}{\beta} \E_\Q \left[\log \frac{d\Q}{d\P} | \mF_t \right]$, where $\Q \in \mathcal{P}$ and $\beta >0$.}  
Then, the one-step risk measure has representation $\rho_t(X_{t+1}) = \esssup_{\Q \in \q_t} \E_\Q[X_{t+1} -\alpha^{\min}_t (\Q)|\mF_t]$, where  $\q_t := \{ \Q \in \mathcal{P} : \Q(B) = \P(B) \text{ for all } B \in \mF_t\}$ is  a set of probability measures that coincide with $\P$ in $\lp{t}$.
If   $\u$ is a static, translation invariant, and positive homogeneous uncertainty set, then the dynamic robust risk measure $R^{\u, \rho}$ admits representation
\begin{align*}\label{eq:R-E-rho-u}
    R_{t,T}^{\u,\rho}(X_{t+1})
    &=
    \esssup \big\{ \rho_t(Z) ~:~  Z \in u_{t+1}(X_{t+1})\big\}
    \\
    &=
    \esssup \Big\{ \esssup_{\Q \in \q_t} \left\{\E_\Q \left[Z- \alpha^{\min}_t (\Q) | \mF_t \right]\right\} ~:~  Z \in u_{t+1}(X_{t+1})\Big\}    
    \\
    &=
    \esssup \big\{ \E\left[Z \tfrac{d\Q}{d\P}  - \alpha^{\min}_t (\Q) | \mF_t \right] ~:~ Z \in u_{t+1}(X_{t+1}), \quad \Q \in \q_t\big\}\\
    &=
    \esssup \big\{ \E[Y | \mF_t] ~:~ Y \in u_{t+1}^*(X_{t+1})\big\}
    \\ 
    &=   
    R_{t,T}^{\u^*,\,\E}(X_{t+1})\,,
\end{align*}
where $\u^*:= \{u^*_t\}_{\tbT}$ is the static uncertainty set given by
\begin{align*}
    u^*_{t+1}(X_{t+1})
    &:=   \bigcup_{\Q \in \q_{t}} u_{t+1}\left(X_{t+1}\frac{d\Q}{d\P} - \alpha^{\min}_t (\Q)  \right)
    \\ &=  \left\{Z \frac{d\Q}{d\P} - \alpha^{\min}_t (\Q) \in \lp{t}~:~ 
    \Q \in \q_{t}^\rho,\,\quad 
    Z \in u_{t+1}(X_{t+1}) 
    \right\}\,.
\end{align*}
This means that any dynamic robust risk measure stemming from a family of convex one-step risk measures can be rewritten as a dynamic robust risk measure, with conditional expectations 
as one-step risk measures and for a suitable dynamic uncertainty set. Thus, the risk $\{\rho_t\}_{\tT}$ can be represented as the uncertainty $\u^*$. However, the converse, i.e. representing the uncertainty as a risk measure, is not necessarily possible, as $u_t(X_t)$ may contain $\mF_t$-measurable random variables that are larger (or smaller) then $\sup X_t$ ($\inf X_t$).

Next, we show that any convex dynamic risk measure can be rewritten as a dynamic robust risk measure, whose one-step risk measures are the conditional expectations. For this, consider the uncertainty sets
     \[ u^{\q_t}_t(X_t) := \left\{ X_t\frac{d\Q}{d\P}  - \alpha^{min}_{t-1} (\Q) : \Q \in \q_t \right\}\,, \qquad \tbT.\]
 Then the dynamic robust risk measure coincides with the dual representation of the convex risk measure used to define the penalty function, i.e., 
 \begin{align*}
        R^{u^{\q_t},\,\E}_{t:T} (X_{t+1}) 
        &=
        \esssup \big\{ \E [ Y | \mF_t  ] : Y \in u^{\q_t}_{t+1}(X_{t+1}) \big\} 
        \\[0.5em]
        &=
        \esssup_{\Q \in \q_t } \big\{ \E_\Q [ X_{t+1}  | \mF_t  ] -\alpha^{min}_t(\Q) \big\} 
        = 
        \rho_t (X_{t+1})\,.
    \end{align*}

\end{example}
}

%
%
%
\begin{APPENDICES}
\section{Additional Proofs.}\label{app:proofs}

The following auxiliary results is need in the proofs that follow.

\begin{lemma}\label{lemma: A=A+A}
    Let $\rho_t$ be normalised, then $\A \subseteq \A + \A $. If, in addition, $\rho_t$ is sub-additive  then $\A = \A + \A$. 
\end{lemma}

\proof{Proof.}
As $\rho_t$ is normalised it holds that $0 \in \A$ and $\A \subseteq \A +  \A$. Next, let $\rho_t$ be sub-additive and $W \in \A + \A$. Then there exists $Y,\,Z \in \A$ such that $W = Y+Z$. By sub-additivity, $\rho_t (W) = \rho_t(Y+Z) \leq \rho_t(Y) + \rho_t(Z) \leq 0  $, and hence, $W \in \A$. 
\hfill \Halmos
\endproof

\proof{Proof of \cref{lemma-u-properties}.} 
Let $X_{t:T}, Y_{t:T} \in \lp{t,T}$ and $B$ be an $\mF_{t-1}$-measurable set. \cref{lemma-u-properties-1}, assume that $u_t$ is normalised and local. Then using locality in the second equation and normalisation in the third, we obtain 
\begin{equation*}
    u_t (1_B \, X_{t:T} ) 
    =
    u_t (1_B \, X_{t:T}  + 1_{B^\comp} \, 0) 
    =
    1_B \, u_t(X_{t:T}) + 1_{B^\comp} \, u_t(0)
    =
    1_B \, u_t(X_{t:T})\,.
\end{equation*}
Next, assume that $u_t \left(1_B \, X_{t:T}\right) = 1_B \, u_t\left(X_{t:T}\right)$ for all $B\in\mF_{t-1}$ and all $X_{t:T}\in\lp{t,T}$. Then by taking $B = \emptyset$, $u_t$ is normalised. We further obtain that $u_t$ is local, by using the assumed property of $U_t$ in the second equation 
\begin{align*}
         u_t (1_B X_{t:T} + 1_{B^\comp} \,Y_{t:T})  
         &= 
         1_B\,  u_t (1_B \,X_{t:T} + 1_{B^\comp} \,Y_{t:T}) + 1_{B^\comp}\, u_t (1_B\, X_{t:T} + 1_{B^\comp} \,Y_{t:T}) 
         \\
         &=
         u_t \left( 1_B \,(1_B \,X_{t:T} + 1_{B^\comp} \,Y_{t:T})\right) 
         + u_t \left( 1_{B^\comp}\,(1_B\, X_{t:T} + 1_{B^\comp}\, Y_{t:T})\right) 
         \\ 
         &=  u_t ( 1_B  \,X_{t:T}  ) + u_t (  1_{B^\comp} \,Y_{t:T}) \,.
\end{align*}

\cref{lemma-u-properties-2}, let $u_t$ be positive homogeneous, then
\begin{equation}\label{pf:eq:pos-hom}
 u_t (1_B \, Z_{t:T} ) = 1_B\,  u_t ( Z_{t:T} ) + 1_{B^\comp}\, u_t (0 )\,, 
 \quad \forall\; Z_{t:T}\in \lp{t,T}\,.   
\end{equation}
Next, we calculate, using \eqref{pf:eq:pos-hom} in the second and forth equation that
\begin{align*}
u_t (1_B X_{t:T} + 1_{B^\comp} Y_{t:T}) + u_t (0) 
         &= 1_B \, u_t (1_B X_{t:T} + 1_{B^\comp} Y_{t:T}) + 1_{B^\comp}\, u_t (1_B X_{t:T} + 1_{B^\comp} Y_{t:T}) + u_t(0)
    \\
         &=  u_t \left(1_B(1_B X_{t:T} + 1_{B^\comp} Y_{t:T})\right) + u_t \left(1_{B^\comp}(1_B X_{t:T} + 1_{B^\comp} Y_{t:T})\right)
    \\
       &=  u_t (1_B\, X_{t:T}) + u_t (1_{B^\comp} Y_{t:T})
         \\
         &= 1_B u_t (X_{t:T})+ 1_{B^\comp}u_t ( Y_{t:T}) +  u_t(0).
 \end{align*}
 Subtracting $u_t(0)$ concludes that $u_t$ is local.
\hfill \Halmos

\proof{Proof of \cref{prop:u_to_R}, Items \ref{u ph - R ph} to \ref{u star - R star}:}

\cref{u ph - R ph}, let $\{\rho_t\}_{\tT}$ and $\u$ be positive homogeneous. 
Then, for all $0 \le \lambda \in \lp{t}$
\begin{align*}
   R_{t,T}( \lambda  X_{t+1:T})
   &= 
   \esssup \left\{ \rho_t (Y) : Y \in u_{t+1} \big(1_{\lambda>0}\, \lambda X_{t+1:T}+1_{\lambda=0}\, \lambda X_{t+1:T}\big)  \right\} 
    \\ 
    &=   
    \esssup \left\{ \rho_t (Y) : Y \in 1_{\lambda>0}\, \lambda u_{t+1} ( X_{t+1:T}) + 1_{\lambda =0}u_{t+1}(0) \right\}
    \\ 
    &=   
    \esssup \left\{ \rho_t (Y + Y') : Y \in 1_{\lambda>0}\, \lambda u_{t+1} ( X_{t+1:T})\,, Y' \in  1_{\lambda =0}u_{t+1}(0) \right\}
    \\
    &=   
    \esssup \left\{ \rho_t (Y + Y') : 1_{\lambda>0}\tfrac{1}{\lambda}\,Y  \in   u_{t+1} ( X_{t+1:T})\,, 1_{\lambda =0}\, Y' \in  u_{t+1}(0) \right\}
    \\
    &=
    \esssup \left\{ \rho_t (1_{\lambda>0} \lambda\,Y + 1_{\lambda=0}\, Y') : Y \in  u_{t+1} ( X_{t+1:T})\,, Y' \in  u_{t+1}(0)  \right\}
    \\
    &=
    \esssup \left\{ 1_{\lambda>0} \lambda\,\rho_t (Y) + 1_{\lambda=0}\, \rho_t (Y') : Y \in  u_{t+1} ( X_{t+1:T})\,, Y' \in  u_{t+1}(0)  \right\}
    \\
    &=
    1_{\lambda>0} \lambda   R_{t,T}( X_{t+1:T}) +
    1_{\lambda=0}   R_{t,T}( 0) 
    .
\end{align*}
If additionally $\u$ is normalised then by \cref{u normalized - R normalize}, $R$ is normalised and the above reduces to $R_{t,T}( \lambda  X_{t+1:T}) =\lambda   R_{t,T}( X_{t+1:T})$ and $R$ is positive homogeneous.

\cref{u conv - R conv}, let $\{\rho_t\}_{\tT}$ be convex and assume for $\tT$ that $u_t(\lambda \, X_{t:T} + (1-\lambda)\, Y_{t:T})\subseteq \lambda \, u_t( X_{t:T}) + (1-\lambda)\, u_t (Y_{t:T})$ for all $\lambda \in \lp{t}$ with $ 0\leq\lambda\leq 1$. Next, define the $\mF_t$-measurable random variables
\begin{equation}\label{eq:indicator-lambda}
 I_0 := \begin{cases}
    0   &  \quad \text{if}\quad \lambda = 0 \\
    \frac{1}{\lambda}  & \quad \text{if} \quad \lambda >0 
\end{cases}     
\qquad \text{and} \qquad
I_1 := \begin{cases}
    0  & \quad  \text{if} \quad \lambda = 1 
    \\ \frac{1}{1-\lambda}  & \quad \text{if} \quad \lambda <1\,.  \end{cases} 
\end{equation}
Then, the robust risk measure satisfies
\begin{align*}
    R_{t,T}( \lambda\,  X_{t+1:T} & +  (1-\lambda) \, Y_{t+1:T} )  
    \\
    &=
    \esssup \left\{ \rho_t (Z) : Z \in u_{t+1} (\lambda \, X_{t+1:T} + (1-\lambda)\,  Y_{t+1:T} ) \right\}
    \\
    &\leq
    \esssup \left\{ \rho_t (Z) : Z \in  \lambda \, u_{t+1}  (X_{t+1:T}) +  (1-\lambda)\,  u_{t+1} (Y_{t+1:T} ) \right\}
    \\
    &=
    \esssup \left\{ \rho_t (  X' + Y' ) : X' \in \lambda \, u_{t+1}  (X_{t+1:T}) , Y' \in  (1-\lambda) \, u_{t+1} (Y_{t+1:T} ) \right\}
    \\
    &=
    \esssup \left\{ \rho_t (  X' + Y' ) : I_0 \, X' \in  u_{t+1}  (X_{t+1:T}) , I_1 \, Y' \in  u_{t+1} (Y_{t+1:T} ) \right\}
    \\
    &=
    \esssup \left\{ \rho_t (  \lambda X' + (1-\lambda)Y' ) : X' \in  u_{t+1}  (X_{t+1:T}) , Y' \in  u_{t+1} (Y_{t+1:T} ) \right\}
    \\
    &\leq
    \esssup \left\{ \lambda\,\rho_t (   X') + (1-\lambda)\,\rho_t(Y' ) : X' \in  u_{t+1}  (X_{t+1:T}) , Y' \in  u_{t+1} (Y_{t+1:T} ) \right\}
    \\ 
    &=
    \lambda \,R_{t,T}(X_{t+1:T}) +  (1-\lambda)\, R_{t,T}(Y_{t+1:T} )\,.
\end{align*}

\cref{u suba- R sub}, 
let $\{\rho_t\}_{\tT}$ be sub-additive and assume that for all $\tbT$ and $Z \in u_t(X_{t:T} + Y_{t:T})$ there exists $X'\in u_t(X_{t:T})$ and $Y' \in u_t(Y_{t:T})$ with $Z \le X' + Y'$. Then
\begin{align*}
    R_{t,T}( X_{t+1:T} &+  Y_{t+1:T} )  
    \\ &= \esssup \left\{ \rho_t (Z) : Z \in u_{t+1} ( X_{t+1:T} +  Y_{t+1:T} ) \right\}
    \\ &\leq \esssup \left\{ \rho_t (  Z) : Z \leq X'+Y' , \; X' \in  u_{t+1}  (X_{t+1:T}) , Y' \in  u_{t+1} (Y_{t+1:T} ) \right\}
\\ &= \esssup \left\{ \rho_t (  X' +Y' ) : X' \in  u_{t+1}  (X_{t+1:T}) , Y' \in  u_{t+1} (Y_{t+1:T} ) \right\}
    \\ &\leq \esssup \left\{   \rho_t (X') +  \rho_t (Y' ) : X' \in  u_{t+1}  (X_{t+1:T}) , Y' \in  u_{t+1} (Y_{t+1:T} ) \right\}
\\ &\leq  R_{t,T}(X_{t+1:T}) +  R_{t,T}(Y_{t+1:T} )\,,
\end{align*}
and $R_{t,T}$ is sub-additive.

 \cref{u sub - R sub}, 
 let $\{\rho_t\}_{\tT}$ be sub-additive and assume that for all $\tbT$, it holds $u_t(X_{t:T} + Y_{t:T}) \subseteq u_t(X_{t:T}) + u_t(Y_{t:T}) $. Then
\begin{align*}
    R_{t,T}( X_{t+1:T} & +  Y_{t+1:T} )  
    \\ &= \esssup \left\{ \rho_t (Z) : Z \in u_{t+1} ( X_{t+1:T} +  Y_{t+1:T} ) \right\}
    \\ 
    &\leq \esssup \left\{ \rho_t (Z) : Z \in  u_{t+1}  (X_{t+1:T}) + u_{t+1} (Y_{t+1:T} ) \right\}
    \\ 
    &= 
    \esssup \left\{ \rho_t (  Z + W ) : Z \in  u_{t+1}  (X_{t+1:T}) , W \in  u_{t+1} (Y_{t+1:T} ) \right\}
    \\
    &\leq
    \esssup \left\{   \rho_t (Z) +  \rho_t (W ) : Z \in  u_{t+1}  (X_{t+1:T}) , W \in  u_{t+1} (Y_{t+1:T} ) \right\}
    \\
    &=
    R_{t,T}(X_{t+1:T}) +  R_{t,T}(Y_{t+1:T} )\,,
\end{align*}
which shows that $R$ is sub-additive.

\cref{u conc - R conc} the proof follows along the exact same steps as in the proof of \cref{u conv - R conv}, with the only difference that the inequalities are in the opposite direction.

\cref{u supe- R sup}, let $\{\rho_t\}_{\tT}$ be super-additive and assume for $\tbT$ and $Z \in u_t(X_{t:T} + Y_{t:T})$ there exist $X' \in u_t(X_{t:T})$ and $Y' \in u_t(Y_{t:T})$ with $Z \ge X' + Y'$. Then
\begin{align*}
    R_{t,T}( X_{t+1:T} & +  Y_{t+1:T} )  
    \\
    &=
    \esssup \left\{ \rho_t (Z) : Z  \in u_{t+1} ( X_{t+1:T} +  Y_{t+1:T} ) \right\}
    \\
    &\ge
    \esssup \left\{ \rho_t (Z) : Z \ge X' + Y', X'\in u_{t+1} ( X_{t+1:T}) \,, Y' \in u_{t+1} (  Y_{t+1:T} ) \right\}
    \\
    &=
    \esssup \left\{ \rho_t (  X' +Y' ) : X' \in  u_{t+1}  (X_{t+1:T}) , Y' \in  u_{t+1} (Y_{t+1:T} ) \right\}
    \\ 
    &\geq 
    \esssup \left\{   \rho_t (X') +  \rho_t (Y' ) : X' \in  u_{t+1}  (X_{t+1:T}) , Y' \in  u_{t+1} (Y_{t+1:T} ) \right\}
    \\ 
    &=  R_{t,T}(X_{t+1:T}) +  R_{t,T}(Y_{t+1:T} )\,,
\end{align*}
and $R_{t,T}$ is super-additive.

\cref{u sup - R sup}, the proof follows along the exact same steps as in the proof of \cref{u sub - R sub}, which the only difference that the inequalities are in the opposite direction.

\cref{u lin - R lin }, this follows from \cref{u sub - R sub} and \cref{u sup - R sup}.

\cref{u star - R star}, the proof follows along the exact same steps as in the proof of \cref{u ph - R ph}, the only difference being that the second equation is an inequality ($\le$) and that $0 \leq \lambda \leq 1$. Further, recall that by definition of $R$, the one-step risk measure $\rho_{t,s}$ is monotone and translation invariance, and thus also local, see e.g., Proposition 3.3 in \cite{cheridito2006dynamic}
\hfill\Halmos
\endproof

\proof{Proof of \cref{theo:equiv_R_U}, Items \ref{theo:equiv_R_U_ph} to \ref{theo:equiv_R_U_star}:}

\cref{theo:equiv_R_U_ph}, let $\rho_t$ be positive homogeneous and $R_{t,T}$ satisfy $ R_{t,T}(\lambda X_{t+1:T}) = \lambda  R_{t,T}( X_{t+1:T}) + 1_{\lambda=0}\, R_{t,T}( 0) $, for all $0 \le \lambda \in \lp{t-1}$. Since $\rho_t$ is positive homogeneous, it holds by Proposition 3.6 in \cite{cheridito2006dynamic} that $\lambda X \in \A$ for all $X \in \A$ and $ 0 \leq \lambda \in \lp{t}$.
This implies that $\A = \lambda \A + 1_{\lambda = 0} \A $ for all $ 0 \leq \lambda \in \lp{t}$.
Next, we obtain, using subsequently  \cref{lemma u to U} \ref{lemma u to U: 4}, the representation of $\A$,
 and finally \cref{lemma u to U} \ref{lemma u to U: 4} that
\begin{align*}\label{equations, proof ph}
    U_{t+1}( \lambda \,X_{t+1:T}) 
    &=
    \A + R_{t,T}(  \lambda\, X_{t+1:T})
    \nonumber \\
    &=
    \A +\lambda \,R_{t,T}( X_{t+1:T}) +  1_{\lambda=0}\, R_{t,T}( 0)
    \\
    &=
    \lambda \,\big( \A + R_{t,T}( X_{t+1:T}) \big)+  1_{\lambda=0}\, \big( \A+R_{t,T}( 0)\big)
   \nonumber \\
    &=
    \lambda \, U_{t+1}(  X_{t+1:T}) +  1_{\lambda=0}\, U_{t+1}( 0)\,, \nonumber
\end{align*}
and $U_{t+1}$ is positive homogeneous.

\cref{theo:equiv_R_U_conv}, let $\{\rho_t\}_{\tT}$ and $R$ be convex and $Y_{t:T} \in \lp{t,T}$. We need to show that for all $\lambda \in \lp{t-1}$ with $0 \le \lambda \le 1$, that 
\begin{equation}\label{eq:U-convex}
     U_{t}(\lambda\, X_{t:T}  +(1-\lambda)\, Y_{t:T})
      \subseteq
     \lambda \, U_{t}(X_{t:T} ) + (1-\lambda)\, U_{t}(Y_{t:T} )\, ,
\end{equation}
which by \cref{lemma u to U} \ref{lemma u to U: 4} is equivalent to 
\begin{align*}
     \A + R_{t,T}(\lambda \, X_{t+1:T} + (1-\lambda)\, Y_{t+1:T})
     &\subseteq
    \lambda \big( \A +  \, R_{t,T}(X_{t+1:T}) \big)+   (1-\lambda)\big( \A + \, R_{t,T}(Y_{t+1:T})\big)
     \\
     &=
     \A + \lambda \, R_{t,T}(X_{t+1:T}) + (1-\lambda)\, R_{t,T}(Y_{t+1:T})
     \,,
\end{align*}
where the equality follows since if $\rho_t$ is convex, then $\A$ is a convex set, hence it holds that $\A = \lambda\A + (1-\lambda )\A$. To show the inclusion, let $Z \in \A + R_{t,T}(\lambda \, X_{t+1:T} + (1-\lambda)\, Y_{t+1:T})$, this means there exists a $Z' \in \lp{t+1}$ with $\rho_t(Z') \le 0$ such that $Z = Z' + R_{t,T}(\lambda \, X_{t+1:T} + (1-\lambda)\, Y_{t+1:T})$. By convexity of $R$ we have
\begin{equation*}
    Z 
    \le
    Z' + \lambda \, R_{t,T}(X_{t+1:T}) + (1-\lambda)\, R_{t,T}(Y_{t+1:T})\,.
\end{equation*}
Therefore, we there exists a $W \ge 0$, such that 
\begin{equation*}
    Z = Z' - W + \lambda \, R_{t,T}(X_{t+1:T}) + (1-\lambda)\, R_{t,T}(Y_{t+1:T})\,.
\end{equation*}
By monotonicity of $\rho_t$ and $W\ge 0$, we have that $\rho_t(Z'-W) \le \rho_t(Z')\le 0$, which implies that $Z \in \A + \lambda\,R_{t,T}( X_{t+1:T}) + (1-\lambda)\, R_{t,T}(Y_{t+1:T})$, and we conclude that \eqref{eq:U-convex} holds.

\cref{theo:equiv_R_U_sub} let $\{\rho_t\}_{\tT}$ and $R$ be sub-additive and $Y_{t+1:T} \in \lp{t+1,T}$. We proceed similarly to the proof of \cref{theo:equiv_R_U_conv}. This means, using \cref{lemma u to U} \ref{lemma u to U: 4}, that we need to show that
\begin{align*}
     \A + R_{t,T}(X_{t+1:T} + Y_{t+1:T})
     &\subseteq
     \A + R_{t,T}(X_{t+1:T}) + \A+ R_{t,T}(Y_{t+1:T})
     \,
     \\ &= \A + R_{t,T}(X_{t+1:T}) + R_{t,T}(Y_{t+1:T})
     \,,
\end{align*}
where the equality follows from sub-additivity of $\rho_t$ and \cref{lemma: A=A+A}.
For the inclusion, let $Z \in \A + R_{t,T}(X_{t+1:T} + Y_{t+1:T})$, this means there exists a $Z' \in \lp{t+1}$ with $\rho_t(Z') \le 0$ such that $Z = Z' + R_{t,T}(X_{t+1:T} + Y_{t+1:T})$. By sub-additivity of $R$ we have
\begin{equation*}
    Z 
    \le
    Z + R_{t,T}(X_{t+1:T}) +  R_{t,T}(Y_{t+1:T})\,,
\end{equation*}
 which by monotonicity of $\rho_t$ implies that $Z \in \A + R_{t,T}(X_{t+1:T}) + R_{t,T}(Y_{t+1:T})$\,.

\cref{theo:equiv_R_U_conc}, let $\{\rho_t\}_{\tT}$ be additive, $R$ concave, $Y_{t+1:T} \in \lp{t+1,T}$, and $\lambda \in \lp{t}$ with $ 0 \leq \lambda \leq 1$. We proceed similar to the proof of \cref{theo:equiv_R_U_conv}. Thus, we need to show that 
\begin{align*}
    \A + \lambda \, R_{t,T}(X_{t+1:T}) + (1-\lambda)\, R_{t,T}(Y_{t+1:T})
     &\subseteq
     \A + R_{t,T}(\lambda \, X_{t+1:T} + (1-\lambda)\, Y_{t+1:T})    
     \,,
\end{align*}
For the inclusion, let $Z \in \A + \lambda \, R_{t,T}(X_{t+1:T}) + (1-\lambda)\, R_{t,T}(Y_{t+1:T})$, this means there exists a $Z' \in \lp{t+1}$ with $\rho_t(Z') \le 0$ such that $Z = Z' + \A + \lambda \, R_{t,T}(X_{t+1:T}) + (1-\lambda)\, R_{t,T}(Y_{t+1:T})$. By sub-additivity of $R$ we have
\begin{equation*}
    Z 
    \le
    Z +  R_{t,T}(\lambda \, X_{t+1:T} + (1-\lambda)\, Y_{t+1:T})  \,,
\end{equation*}
 which by monotonicity of $\rho_t$ implies that $Z \in \A + R_{t,T}(\lambda \, X_{t+1:T} + (1-\lambda)\, Y_{t+1:T})  $\,.

\cref{theo:equiv_R_U_sup}, the proof follows using similar steps as in the proof of \cref{theo:equiv_R_U_sup}.

\cref{theo:equiv_R_U_lin} is a consequence of \cref{theo:equiv_R_U_sup,theo:equiv_R_U_sub}.

 \cref{theo:equiv_R_U_star} follows the same steps of the proof of \cref{theo:equiv_R_U_ph}, where the second equality becomes a set inclusion, i.e. $\A + R_{t,T}(  \lambda\, X_{t+1:T}) \subseteq \A +\lambda \,R_{t,T}( X_{t+1:T}) +  1_{\lambda=0}\, R_{t,T}( 0)$. This follows from $R_{t,T}(  \lambda\, X_{t+1:T})  \leq \lambda R_{t,T}( X_{t+1:T}) +  1_{\lambda=0}\, R_{t,T}( 0)$ and monotonicity of the one-step risk measure $\rho_t$.
\hfill \Halmos
\endproof

\end{APPENDICES}

\section*{Acknowledgements.}
 M. Mailhot acknowledges the support from the Natural Sciences and Engineering Research Council of Canada (grant RGPIN-2015-05447). S. Pesenti acknowledges the support from the Natural Sciences and Engineering Research Council of Canada (grants DGECR-2020-00333 and RGPIN-2020-04289). M. Mailhot and S. Pesenti also acknowledge the support from the Canadian Statistical Sciences Institute (CANSSI).


\bibliographystyle{informs2014} 
\bibliography{references} 


\end{document}